\documentclass{article}
\usepackage[totalwidth=480pt, totalheight=680pt]{geometry}

\newif\ifARXIV
\ARXIVtrue

\usepackage{url}
\usepackage{algorithm}
\usepackage{algpseudocode}
\usepackage{verbatim}
\usepackage{color}

\usepackage{amsmath,amsthm}
\usepackage{amssymb}
\usepackage{amsfonts}
\usepackage{latexsym}
\usepackage{stmaryrd}
\usepackage{listings}
\usepackage{enumerate}
\usepackage{xspace}
\usepackage{subfigure}
\usepackage{multicol}
\usepackage{tikz}
\usetikzlibrary{calc,shapes,arrows}
\usepackage{pbox}
\usepackage{graphicx}
\usepackage{proof}
\usepackage{xspace} 
\usepackage{amsthm}
\usepackage{multirow}
\usepackage{hhline}
\usepackage{array}
\usepackage{balance}
\usepackage{ifthen}
\usepackage{enumitem}
\usepackage{mathtools}
\usepackage{balance}

\usepackage{enumitem}
\setlist{parsep=0pt,listparindent=\parindent}

\usepackage[toc,page]{appendix}

\definecolor{light-gray}{gray}{0.80}
\usepackage[
  draft,
  colorlinks,
  urlcolor=blue,
  citecolor=blue,
  pdfauthor={Zvonimir Pavlinovic and Tim King and Thomas Wies},
]{hyperref}

\newcommand{\smartparagraph}[1]{\smallskip\noindent 
{\bf #1}\ }

\newenvironment{proofsketch}{\noindent {\em Proof Sketch.}}{\mbox{} \hfill $\Box$ \vspace{2ex} }

\newcommand{\minerrors}{\textsc{Solve}}
\newcommand{\usages}{\mathit{Usages}}
\newcommand{\pdefs}{\mathit{PDefs}}
\newcommand{\pdef}{\mathit{PDef}}
\newcommand{\scope}{\mathit{Scope}}

\newcommand{\Relax}{\ensuremath{\mathrm{relax}}}
\newcommand{\Clause}{\ensuremath{C}}
\newcommand{\Clauses}{\mathcal{C}}
\newcommand{\HClauses}{\Clauses_H}
\newcommand{\SClauses}{\Clauses_S}
\newcommand{\RClauses}{\SClauses^{\Relax}}

\newcommand{\Assumptions}{\mathcal{A}}

\newcommand{\Th}{\mathcal{T}}
\DeclareMathOperator{\modelsTh}{\models_{\Th}}
\newcommand{\Assertions}{\Phi}
\newcommand{\assertion}{\varphi}
\newcommand{\HAssertions}{\Assertions_H}
\newcommand{\SAssertions}{\Assertions_S}

\DeclareMathOperator{\Imp}{\Rightarrow}

\newcommand{\Selector}{s}
\newcommand{\Selectors}{\mathcal{S}}

\newcommand{\Principal}{\rho}
\newcommand{\UsePrincipal}[1]{P_{#1}}

\newcommand{\Assignment}{M}
\newcommand{\True}{\texttt{true}}
\newcommand{\False}{\texttt{false}}
\DeclareMathOperator*{\Maximize}{maximize}

\newcommand{\Problem}[1]{\mathbf{#1}}
\newcommand{\Weight}{\ensuremath{w}}

\newcommand{\Lemmas}{\mathcal{L}}

\newcommand{\Unsat}{\mathbf{unsat}\xspace}

\newcommand{\tbool}{\mathsf{bool}}
\newcommand{\tint}{\mathsf{int}}

\newcommand{\tstring}{\mathsf{string}}
\newcommand{\funconstr}{\mathsf{fun}}
\newcommand{\tripleconstr}{\mathsf{triple}}
\newcommand{\fun}[2]{\funconstr(#1, #2)}
\newcommand{\triple}[3]{\tripleconstr(#1, #2, #3)}

\newcommand{\tvar}{\alpha}

\newcommand{\hole}[0]{\bot}
\newcommand{\lang}[0]{$\lambda^{\bot}$\xspace}

\newcommand{\Let}{\mathtt{let}}
\newcommand{\letin}[3]{\Let\; #1 = #2 \; \mathtt{in}\; #3}
\newcommand{\ite}[3]{\mathtt{if}\; #1 \; \mathtt{then} \; #2 \;
\mathtt{else} \; #3}
\newcommand{\lm}[2]{\lambda #1.\, #2}

\newcommand{\prog}{p}
\newcommand{\locs}{\mathit{Loc}}

\newcommand{\explocs}{\mathfrak{L}}
\newcommand{\trel}[1]{\vdash_{#1}}
\newcommand{\diffset}{\mathfrak{D}}

\newcommand{\mask}{\mathit{mask}}
\newcommand{\BB}{\mathbb{B}}
\newcommand{\ZZ}{\mathbb{Z}}
\newcommand{\NN}{\mathbb{N}}
\newcommand{\fv}{\mathit{fv}}

\newcommand{\pto}{\rightharpoonup}
\newcommand{\dom}{\mathsf{dom}}

\newcommand{\pset}[2]{\left\{\,#1\mid#2\,\right\}}

\newcommand{\prop}[0]{T\xspace}
\newcommand{\instance}{\mathit{I}}

\newtheorem{definition}{Definition}
\newtheorem{lemma}{Lemma}
\newtheorem{theorem}{Theorem}

\newcommand{\Dlocs}{\mathit{dloc}}
\newcommand{\Ulocs}{\mathit{Uloc}}
\newcommand{\Vlocs}{\mathit{Vloc}}

\newcommand{\stringquote}[1]{\texttt{"} #1 \texttt{"}}

\usepackage{authblk}

\title{On Practical SMT-Based Type Error Localization}
\author[1]{Zvonimir Pavlinovic}
\author[2]{Tim King}
\author[1]{Thomas Wies}
\affil[1]{New York University}
\affil[2]{Verimag}

\begin{document}

\maketitle

\begin{abstract} 
  Compilers for statically typed functional programming languages are
  notorious for generating confusing type error messages. When the
  compiler detects a type error, it typically reports the program
  location where the type checking failed as the source of the
  error. Since other error sources are not even considered, the actual
  root cause is often missed. A more adequate approach is to consider
  all possible error sources and report the most useful one subject to
  some usefulness criterion. In our previous work, we showed that this
  approach can be formulated as an optimization problem related to
  satisfiability modulo theories (SMT). This formulation cleanly
  separates the heuristic nature of usefulness criteria from the
  underlying search problem. Unfortunately, algorithms that search for
  an optimal error source cannot directly use principal types which
  are crucial for dealing with the exponential-time complexity of the
  decision problem of polymorphic type checking. In this paper, we
  present a new algorithm that efficiently finds an optimal error
  source in a given ill-typed program. Our algorithm uses an improved
  SMT encoding to cope with the high complexity of polymorphic typing
  by iteratively expanding the typing constraints from which principal
  types are derived. The algorithm preserves the clean separation
  between the heuristics and the actual search.  We have implemented
  our algorithm for OCaml.  In our experimental evaluation, we found
  that the algorithm reduces the running times for optimal type error
  localization from minutes to seconds and scales better than previous
  localization algorithms.
\end{abstract}






\section{Introduction}
\label{sec:introduction}

Hindley-Milner type systems support automatic type inference, which is
one of the features that make languages such as Haskell, OCaml, and
SML so attractive. While the type inference problem for these
languages is well understood~\cite{principal, mldexp1, mldexp2, hmx,
  aiken, ctypes}, the problem of diagnosing type errors still lacks
satisfactory
solutions~\cite{wand, explaining, slicing, flow, comp, counter,
  discriminative, seminal, minerrors}.

When type inference fails, a compiler usually reports the location
where the first type mismatch occurred as the source of the error.
However, often the actual location that is to blame for the error and
needs to be fixed is somewhere else entirely. Consequently, the
quality of type error messages suffers, which increases the debugging
time for the programmer. A more adequate approach is to consider all
possible error sources and then choose the one that is most likely to
blame for the error. Here, an error source is a set of program
locations that, once corrected, yield a well-typed program.

The challenge for this approach is that it involves two
subproblems that are difficult to untangle: (1) searching for type
error sources, and (2) ranking error sources according to some
usefulness criterion (e.g., the number of required modifications to
fix the program). Existing solutions to type error localization make
specific heuristic decisions for solving these subproblems. As a
consequence, the resulting algorithms often do not provide formal
guarantees or use specific usefulness criteria that are difficult to
justify or adapt. In our recent work~\cite{minerrors}, we have
proposed a novel approach that formalizes type error localization as
an optimization problem. The advantage of this approach is that it
creates a clean separation between (1) the algorithmic problem of
finding error sources of minimum cost, and (2) the problem of finding
good usefulness criteria that define the cost function. This
separation of concerns allows us to study these two problems
independently. In this paper, we develop an efficient solution for
problem (1).

\smartparagraph{Challenge.} Type inference is often formalized in
terms of constraint satisfaction~\cite{hmx, aiken, ctypes}. In this
formalization, each expression in the program is associated with a
type variable. A typing constraint of a program encodes the
relationship between the type of each expression and the types of its
subexpressions by constraining the type variables appropriately. The
program is then well-typed iff there exists an assignment of types to
the type variables that satisfies the constraint. In our previous
paper, we used this formalization to reduce the problem of finding
minimum error sources to a known optimization problem in
satisfiability modulo theories (SMT), the partial weighted MaxSMT
problem. This reduction enables us to use existing MaxSMT solvers for
type error localization.

The reduction to constraint satisfaction also has its problems.  The
number of typing constraints can grow exponentially in the size of the
program.  This is because the constraints associated with polymorphic
functions are duplicated each time these functions are used. This
explosion in the constraint size does not seem to be avoidable because
the type inference problem is known to be
EXPTIME-complete~\cite{mldexp1, mldexp2}. However, in practice,
compilers successfully avoid the explosion by computing the principal
type~\cite{principal} of each polymorphic function and then
instantiating a fresh copy of this type for each usage. The resulting
constraints are much smaller in practice. Since the smaller
constraints are equisatisfiable with the original constraints, the
resulting algorithm is a decision procedure for the type checking
problem~\cite{principal}. Unfortunately, this technique cannot be
applied immediately to the optimization problem of type error
localization. If the minimum cost error source is located inside of a
polymorphic function, then abstracting the constraints of that
function by its principle type will hide this error source.  Thus,
this approach can yield incorrect results.  This dilemma is inherent to all
type error localization techniques and the main reason why existing
algorithms that are guaranteed to produce optimal solutions do not yet
scale to real-world programs.


\smartparagraph{Solution.} 
Our new algorithm makes the optimistic assumption that the relevant
type error sources only involve few polymorphic functions, even for
large programs. Based on this assumption, we propose an improved
reduction to the MaxSMT problem that abstracts polymorphic functions
by principal types. The abstraction is done in such a way that all
potential error sources involving the definition of an abstracted
function are represented by a single error source whose cost is
smaller or equal to the cost of all these potential error sources.
The algorithm then iteratively computes minimum error sources for
abstracted constraints. If an error source involves a usage of a
polymorphic function, the corresponding instantiations of the
principal type of that function are expanded to the actual typing
constraints. Usages of polymorphic functions that are exposed by the
new constraints are expanded if they are relevant for the minimum
error source in the next iteration. The algorithm eventually
terminates when the computed minimum error source no longer involves
any usages of abstracted polymorphic functions. Such error sources are
guaranteed to have minimum cost for the fully expanded constraints,
even if the final constraint is not yet fully expanded.

We have implemented our algorithm targeting OCaml~\cite{ocaml} and
evaluated it on benchmarks for type error localization~\cite{seminal}
as well as code taken from a larger OCaml application.  We used
EasyOCaml~\cite{ecaml} for generating typing constraints and the
MaxSMT solver $\nu$Z~\cite{nuZ,newZ,z3} for computing minimum error
sources. We found that our implementation efficiently computes the
minimum error source in our experiments for a typical usefulness
criterion taken from~\cite{minerrors}. In particular, our algorithm is
able to compute minimum error sources for realistic programs in
seconds, compared to several minutes for the naive algorithm and
other approaches. Also, on our benchmarks, the new algorithm avoids
the exponential explosion in the size of the generated constraints
that we observe in the naive algorithm.





\smartparagraph{Related Work.}  
The formulation of type error localization as an optimization problem
follows our previous work~\cite{minerrors}. There, we presented the
naive implementation of the search algorithm. Other work on type error
localization is not directly comparable to ours. Most closely related
is the work by Zhang and Myers~\cite{myers,zhangPLDI} where type error
localization is cast as a graph analysis problem. Their approach,
however, does not address the issue of constraint explosion, which
here manifests as an explosion in the size of the generated
graphs. In fact, our algorithm is faster than their implementation on the
same benchmarks: while their tool runs over a minute for some programs,
our algorithm always finishes in a just of a couple of seconds. Consequently,
for larger problem instances with a couple of thousands of lines of code their
implementation runs out of memory. Our algorithm, on the other hand, finishes
in less than 50 seconds. The majority of the remaining work on type error localization
is concerned with different definitions and notions of usefulness
criteria~\cite{wand,explaining,slicing,flow,comp,counter,discriminative}.
In our previous work, we gave experimental evidence that our approach
yields better error sources than the OCaml compiler even for a
relatively simple cost function. The work in this paper is orthogonal
because it focuses on practical algorithms for computing a minimum
error source subject to an arbitrary cost function.

\smartparagraph{Contributions.}  Our contributions can be summarized
as follows:

\begin{itemize}
\item We present a new algorithm that uses SMT techniques to
  efficiently find the minimum error source in a given ill-typed
  program. The algorithm works for an arbitrary cost function which
  encodes the usefulness criterion for ranking error sources.

\item We have implemented the algorithm and showed that it scales to
  programs of realistic size.

\item To our knowledge, this is the first algorithm for type error
  localization that gives formal optimality guarantees and has the
  potential to be usable in practice.

\end{itemize}


\section{Overview}
\label{sec:overview}

In this section we provide an overview of our approach through an illustrative 
example. We start by describing type error localization as an optimization problem 
and then exemplify the workings of our algorithm that efficiently
solves the problem. 

\subsection{Example}

Our running OCaml example is as follows:

\begin{center}
\begin{lstlisting}
let first  (a, b, _) = a @\label{code:first-def-loc}@
let second (a, b, _) = b @\label{code:second-def-loc}@
let f x = @\label{code:f-def-loc}@
  let first_x  = first x in @\label{code:wrong-call}@
  let second_x = int_of_string (second x) in@\label{code:ok-call}@
  first_x + second_x
f ("1", "2", f ("3", "4", 5)) @\label{code:error}@
\end{lstlisting}
\end{center}


\noindent This program is not well-typed. While polymorphic functions \texttt{first}, 
\texttt{second}, and \texttt{f} do not have any type errors,
the calls to \texttt{f} on line~\ref{code:error} are ill-typed.
The inner call to \texttt{f} is passed a triple having the string \texttt{"3"}
as its first member, whereas an integer is expected.
The standard OCaml compiler~\cite{ocaml} reports this type error to the
programmer blaming expression \texttt{"1"} on line~\ref{code:error} as the source of the 
error (OCaml version $4.01.0$). However, perhaps the programmer made a mistake by calling function \texttt{first} 
on line~\ref{code:wrong-call} or maybe she incorrectly defined \texttt{first} on 
line~\ref{code:first-def-loc}. 
Maybe the programmer should have wrapped this call with a call
to \texttt{int\_of\_string} just as she has done on line~\ref{code:ok-call}.
The OCaml compiler disregards such error sources. 

\subsection{Finding Minimum Error Sources}
In our previous work~\cite{minerrors}, we formulated type error localization as an optimization
problem of finding an error source that is considered most useful for the programmer. 
The criterion for usefulness is provided by the compiler. We define an error source to be a set 
of program expressions that, once fixed, make the program well-typed.
A usefulness criterion is a function from program expressions to positive weights.
A minimum error source is an error source with minimum cumulative weight.
It corresponds to the most useful error source. To make this more clear,
consider a usefulness criterion where each expression is assigned a weight equal
to the size of the expression, represented as an abstract syntax tree (AST).
In the example, expression 
\texttt{first} on line~\ref{code:wrong-call} is a singleton error source of
weight $1$ as replacing it by a function of a type that is an instance of the polymorphic type
\[ \forall \tvar. \fun{\tstring * \tstring * \tvar}{\tint},\] makes
the program well-typed, say
\texttt{int\_of\_string}$\;\circ\;$\texttt{first}. Similarly,
replacing the expression \texttt{a} on line~\ref{code:first-def-loc}
with \texttt{(int\_of\_string a)} also resolves the type error.
Loosely speaking, the error sources that are minimum subject to the
AST size criterion require the fewest corrections to fix the error.
The two error sources described above are minimum error sources since
their cumulative weight is $1$, which is minimum for this program and
criterion.  In contrast, we could abstract the entire application
\texttt{first x} on the same line to get a well typed program. Thus
\texttt{first x} is also an error source, but it is not minimum as its
weight is $3$ according to its AST size (\texttt{first}, \texttt{x},
and function application).  Note that \texttt{"1"} on line
\ref{code:error} on its own is not an error source according to our
definition.  If one abstracts \texttt{"1"}, this does not yield a well
typed program since the expression \texttt{"3"} on line
\ref{code:error} would still lead to a failure.  Abstracting both
$\{\texttt{"1"}, \texttt{"3"}\}$ is an error source with cumulative
weight 2.  Observe that there is a clean separation between searching
for a minimum error source and the definition of the usefulness
criterion.  This allows easy prototyping of various criteria without
modifying the compiler infrastructure.  A more detailed discussion of
potential usefulness criteria can be found in~\cite{minerrors}.

\subsection{Abstraction  by Principal Types}

A potential obstacle to adopting this approach is that compilers now
need to solve an optimization problem instead of a decision problem.
This is particularly problematic since type checking for polymorphic
type systems is \textmd{EXPTIME} complete~\cite{mldexp1,
  mldexp2}. This high complexity manifests in an exponential number of
generated typing constraints. For instance, consider the typing
constraints for the function \texttt{second}:

\begin{align}
  \tvar_{\mathit{second}} & = \fun{\tvar_i}{\tvar_o} & \text{[Def.\ of \lstinline!second!]}\label{eq:seconddef} \\
  \tvar_i & = \triple{\tvar_a}{\tvar_b}{\tvar_{\_}} & \text{\lstinline!(a, b, _)!} \label{eq:abc}\\
  \tvar_o & = \tvar_b & \text{\lstinline!b!} \label{eq:b}
\end{align}

\noindent The above constraints state that the type of
\texttt{second}, represented by the type variable $\tvar_{second}$, is
a function type (1) that accepts some triple (2) and returns a value
whose type is equal to the type of the second component of that triple
(3). When a polymorphic function, such as \texttt{second}, is called
in the program, the associated set of typing constraints needs to be
\textit{instantiated} and the new copy has to be added to the whole
set of typing constraints. Instantiation of typing constraints
involves copying the constraints and replacing free type variables in
the copy with fresh type variables. In our example, each call to
\texttt{second} in \texttt{f} is accounted for by a fresh instance of
$\tvar_{second}$ and the whole set of associated typing constraints is
copied and instantiated by replacing the type variable
$\tvar_{second}$ with a fresh type variable.  If the constraints of
polymorphic function were not freshly instantiated for each usage of
the function, the same type variable would be constrained by the
context of each usage, potentially resulting in a spurious type error.

Instantiation of typing constraints as described above leads to an
explosion in the total number of generated constraints. For instance,
the typing constraints for each call to \texttt{f} are instantiated
twice.  Each of these copies in turn includes a fresh copy of the
constraints associated with each call to \texttt{second} and
\texttt{first} in \texttt{f}.  Hence, the number of typing constraints
can grow exponentially, to the point where the whole approach becomes
impractical.  To alleviate this problem, compilers first solve the
typing constraints for each polymorphic function to get their
principal types. Intuitively, the principal type is the most general
type of an expression~\cite{principal}.  Then, each time the function
is used only its principal type is instantiated, instead of the whole
set of associated typing constraints.

In the example, when typing the line \ref{code:error}, the typing environment
contains principal types for \texttt{first}, \texttt{second}, and \texttt{f}
(given as comments below).
\begin{center}
\begin{lstlisting}[mathescape]
; first  : $\forall \tvar_a, \tvar_b, \tvar_c. \;\fun{\tvar_a * \tvar_b
  * \tvar_c}{\tvar_a}$
; second : $\forall \tvar_a, \tvar_b, \tvar_c. \;\fun{\tvar_a * \tvar_b
  * \tvar_c}{\tvar_b}$
; f : $\forall \tvar_a. \; \fun{\tint * \tstring * \tvar_a}{\tint}$
f ("1", "2", f ("3", "4", 5))
\end{lstlisting}
\end{center}
The bodies of the three bound variables and the typing constraints within the
bodies are effectively abstracted at this point.
Type inference instantiates the principal type of \texttt{f},
\[ \texttt{f} :  \forall \tvar. \fun{\tint * \tstring * \tvar}{\tint}, \]
but this will fail to unify with the argument to \texttt{f}
which has type \texttt{string * string * int}.

The principal type technique for avoiding the constraint explosion
works very well in practice for the decision problem of type checking.
However, we will need to adapt it in order to work with the
optimization problem of searching for a minimum error source.  When
the search algorithm checks whether a set of expressions is an error
source, it checks satisfiability of the typing constraints that have
been generated for the whole program, where the constraints for the
expressions in the potential error source have been removed. If we
directly use the principal type as an abstraction of the function
body, we potentially miss some error sources that involve expressions
in the abstracted function body.  To illustrate this point, consider
the principal type abstraction of our example program above. The
application of the expression \texttt{first} at line
\ref{code:wrong-call} has in effect been abstracted from the program
and cannot be reported as an error source, although it is in fact
minimum. In general, fixing an error source in a function definition
can change the principal type of that function. The search algorithm
must take such changes into account in order to identify the minimum
error sources correctly. In our running example, a generic fix to the
call to function \texttt{first} at line \ref{code:wrong-call} results
in the principal type of \texttt{f} being:
\[ \forall \tvar_a, \tvar_b. \fun{\tvar_a * \tstring * \tvar_b}{\tint}. \]
Additionally, principal types may not exist for some expressions in an
ill-typed program. The algorithm needs to handle such cases gracefully.

\subsection{Approach}

Our solution to this problem is an algorithm that finds a minimum
error source by expanding the principal types of polymorphic functions
iteratively.  We first compute principal types for each let-bound
variable whenever possible.  We begin our search assuming that none of
the usages of the variables whose principal types could be computed are
involved in a minimum error source. Each principal type is assigned
the minimum weight of all constraints in the associated let
definition, conservatively approximating the potential minimum
error sources that involve these constraints.

\begin{figure}
\begin{equation*}
  \begin{array}{lr}
  \UsePrincipal{\texttt{first}} & 1 \\
  \UsePrincipal{\texttt{second}} & 1 \\
  \UsePrincipal{\texttt{f}} & 1 \\
  \tvar_{f1} = \fun{\tvar_{i1}}{\tvar_{o1}} & 11 \\
  \UsePrincipal{\texttt{f}} \Rightarrow \tvar_{f1} = \fun{\tint * \tstring * \tvar'}{\tint} & 1 \\
  \tvar_{i1} = \tvar_{\stringquote{1}} * \tvar_{\stringquote{2}} * \tvar_{app} & 9 \\
  \tvar_{\stringquote{1}} = \tstring & 1\\
  \tvar_{\stringquote{2}} = \tstring & 1\\
  \tvar_{app} = \tvar_{o2} & 6\\
  \tvar_{f2} = \fun{\tvar_{i2}}{\tvar_{o2}} & 6 \\
  \UsePrincipal{\texttt{f}} \Rightarrow \tvar_{f2} = \fun{\tint * \tstring * \tvar''}{\tint} & 1 \\
  \tvar_{i2} = \tvar_{\stringquote{4}} * \tvar_{\stringquote{5}} * \tvar_{6} & 4 \\
  \tvar_{\stringquote{4}} = \tstring & 1\\
  \tvar_{\stringquote{5}} = \tstring & 1\\
  \tvar_{6} = \tint & 1 \\
  \end{array}
  \label{eqn:ov:example}
\end{equation*}
\caption{
  Typing constraints and weights for the first iteration of the
  localization algorithm.
}\label{fig:typing:example}
\end{figure}

In our example, this results in exactly the same abstraction of the program
as before, and the weights of \texttt{f}, \texttt{first} and \texttt{second}
are all $1$.
We write the proposition that the principle type for \texttt{foo} is correct as $\UsePrincipal{\mathit{foo}}$.
Typing for each call to \texttt{f} is represented with a fresh instance of the
corresponding principal type.
Each usage of \texttt{f} is marked as depending on the principal type for \texttt{f},
and is guarded by $\UsePrincipal{f}$.
Figure~\ref{fig:typing:example} gives the typing constraints and the weight of each constraint.\footnote{
  This is slightly simplified from the actual encoding in \S\ref{sec:algorithm}.
}

The above set of constraints is unsatisfiable.  The minimum error
source for these constraints is to relax the constraint for
$\UsePrincipal{\texttt{f}}$.  This indicates that we cannot rely on
the principal type for \texttt{f} to find the minimum error source
for the program.  We relax the assumption that
$\UsePrincipal{\texttt{f}}$ is true, and include the body for
\texttt{f} in our next iteration.  This next iteration is effectively
analyzing the program:
\begin{center}
\begin{lstlisting}[mathescape]
; first  : $\forall \tvar_a, \tvar_b, \tvar_c. \;\fun{\tvar_a * \tvar_b
  * \tvar_c}{\tvar_a}$
; second : $\forall \tvar_a, \tvar_b, \tvar_c. \;\fun{\tvar_a * \tvar_b
  * \tvar_c}{\tvar_b}$
let f x =
  let first_x  = first x in
  let second_x = int_of_string (second x) in
  first_x + second_x
f ("1", "2", f ("3", "4", 5))
\end{lstlisting}
\end{center}
\noindent Here, typing for each usage of \texttt{first} and
\texttt{second} is represented by fresh instances of the corresponding
principal types.  As in the previous iteration, we again compute a
minimum error source and decide whether further expansions are
necessary.  In the next iteration, the unique minimum error source of
the new abstraction is the application of \texttt{first} on
line~\ref{code:wrong-call}, which is also a minimum error source of
the whole program. Note that this new minimum error source does not
involve any expressions with unexpanded principal types. Hence, we can
conclude that we have found a true minimum error source and our
algorithm terminates. That is, the algorithm stops before the
principal types for \texttt{second} and \texttt{first} have been
expanded. The procedure only expands the usages of those polymorphic
functions that are involved in the error when necessary, thus lazily
avoiding the constraint explosion.  This is sound because of the
conservative abstraction of potential error sources in the unexpanded
definitions. In our running example, we can conclude from the
constraints of the final iteration that fixing \texttt{second} does
not resolve the error and fixing \texttt{first} is not cheaper than
just fixing the call to \texttt{first} ($P_{first}$ is an error source
of weight $1$ but so is the call to \texttt{first}). Hence, the
algorithm yields a correct result.

The search for a minimum error source in each iteration is performed
by a weighted partial MaxSMT solver.  In Section~\ref{sec:problem}, we provide the formal
definitions of the problem of finding minimum error sources and the
weighted partial MaxSMT problem. Section~\ref{sec:algorithm} describes
the iterative algorithm that reduces the former problem to the later
and argues its correctness. In Section~\ref{sec:evaluation}, we
present our experimental evaluation.
\ifARXIV
\else
The details of the correctness
proof can be found in Appendix~\ref{sec:appendix-l2}.
\fi

\section{Background}
\label{sec:problem}
We recall in this section the  minimum error source~\S\ref{sec:minerror} problem
from~\cite{minerrors}
as well as our targeted language~\S\ref{sec:lang} and type system~\S\ref{sec:types}.
We also describe the satisfiability~\S\ref{sec:sat},
MaxSAT~\S\ref{sec:maxsat}, and
MaxSMT~\S\ref{sec:maxsmt} problems
used to solve the minimum error source problem in~\S\ref{sec:algorithm}.

\subsection{Language}
\label{sec:lang}
Our presentation is based on an idealized lambda calculus, called  \lang, with let
polymorphism, conditional branching, and special value $\hole$ called
\emph{hole}.
Holes allow us to create expressions that have the most general type
(\S\ref{sec:types}).
\begin{align*}
  \textbf{Expressions} \qquad e ::= \; & x & \text{variable} \\
  \mid \; & v & \text{value} \\
  \mid \; & e \; e & \text{application} \\
  \mid \; & \ite{e}{e}{e} & \text{conditional} \\
  \mid \; & \letin{x}{e}{e} & \text{let binding} \\[0.15cm]
  \textbf{Values} \qquad v ::= \; & n & \text{integers} \\
  \mid \; & b & \text{Booleans} \\
  \mid \; & \lm{x}{e} & \text{abstraction} \\
  \mid \; & \hole & \text{hole}
\end{align*}
Values in the language include integer constants, $n \in \ZZ$,
Boolean constants, $b \in \BB$, and lambda abstractions.
The let bindings allow for the definition of polymorphic functions.
We assume an infinite set of program variables, $x, y, \ldots$.
Programs are expressions in which no variable is free.
The reader may assume the expected semantics (with $\hole$ acting
as an exception).

\subsection{Types}
\label{sec:types}
Every type in \lang is a monotype or a polytype.
\begin{align*}
  \textbf{Monotypes} \qquad \tau ::= \; & \tbool  \mid \tint \mid 
  \tvar \mid \tau \rightarrow \tau \\
  \textbf{Polytypes} \qquad \sigma ::= \; & \tau \mid \forall \tvar. \sigma
\end{align*}
A monotype $\tau$ is either a \emph{base type} $\tbool$ or $\tint$,
a \emph{type variable} $\tvar$, or a \emph{function type} $\tau \rightarrow \tau$.
The \emph{ground} types are monotypes in which no type variable occurs.

A polytype is either a monotype or the quantification of a type variable
over a polytype.
A polytype $\sigma$ can always be written $\forall \tvar_1. \cdots \forall \tvar_n. \tau$
where $\tau$ is a monotype or in shorthand, $\forall \vec{\tvar}. \tau$.
The set of free type variables in $\sigma$ is denoted $\fv(\sigma)$.
We write $\sigma[\tau_1/\tvar_1, \ldots, \tau_n/\tvar_n]$ for capture-avoiding
substitution in $\sigma$ of free occurrences of the type variable
$\tvar_i$ by the monotype $\tau_i$.
We uniformly shorten this to $\sigma[\tau_i/\tvar_i]$ to denote $n$-ary substitution.
The polytype $\forall \vec{\tvar}. \tau$ is considered to represent
all types obtained by instantiating
the type variables $\vec{\tvar}$ by ground monotypes,
e.g. $\tau[\tau_i/\tvar_i]$.
Finally, the polytype $\sigma = \forall \vec{\alpha}. \tau$ has a generic
instance $\sigma' = \forall \vec{\beta}. \tau'$ if
$\tau' = \tau[\tau_i/\alpha_i]$ for some
monotypes $\tau_1, \ldots, \tau_n$ and $\vec{\beta} \not\in \fv(\sigma)$.

Like other Hindley-Milner type systems, type inference is decidable for \lang.
A \emph{typing environment} $\Gamma$ is a mapping of variables to types.
We denote by $\Gamma \vdash e: \tau$ the typing judgment
that the expression $e$ has type $\tau$ under a typing environment $\Gamma$.
The free variables of $\Gamma$ are denoted as $\fv(\Gamma)$.
A program $\prog$ is \emph{well typed} iff
the empty typing environment $\emptyset$ can infer a type for $\prog$,
$\emptyset \vdash \prog:\sigma$.

Figure~\ref{fig:typing} gives the typing rules for \lang.
The \textsc{[Hole]} rule is non-standard and
states that the expression $\hole$ has the polytype $\forall \alpha. \alpha$.
During type inference, the rule \textsc{[Hole]} assigns to each usage of $\hole$ a
fresh unconstrained type variable.
Hole values may always safely be used without causing a type error.
We may think of $\hole$ in two ways:
as exceptions in OCaml~\cite{seminal}, or
as a place holder for another expression.
In \S\ref{sec:minerror}, we abstract sub-expressions
in a program $\prog$ as $\hole$ to obtain a new program $\prog'$ that is well typed.

\ifARXIV
\begin{figure*}[t]
  \centering
  \begin{gather*}
    \begin{array}{cc}
      \infer[\textsc{[Abs]}] {\Gamma \vdash \lambda x.e:\tau_1 \rightarrow \tau_2}{\Gamma.x:\tau_1 \vdash e:\tau_2} \qquad&
      \infer[\textsc{[App]}] {\Gamma \vdash  e_1 \; e_2:\tau_2}{\Gamma \vdash e_1:\tau_1 \rightarrow \tau_2 \quad \Gamma \vdash e_2:\tau_1}
    \end{array} \\[0.8cm]      
    \infer[\textsc{[Cond]}] {\Gamma \vdash \ite{e_1}{e_2}{e_3}:\tau}{\Gamma \vdash e_1:\tbool \quad \Gamma \vdash e_2:\tau \quad \Gamma \vdash e_3:\tau}
    \\[0.8cm]
    \begin{array}{ccc}
      \infer[\textsc{[Hole]}] {\Gamma \vdash \hole:\alpha}{\alpha\; \text{new}} \qquad&
      \infer[\textsc{[Bool]}] {\Gamma \vdash b:\tbool}{b \in \mathbb{B}} \qquad&
      \infer[\textsc{[Int]}] {\Gamma \vdash n:\tint}{n \in \mathbb{Z}}
    \end{array}
    \\[0.8cm]
    \begin{array}{cc}
      \infer[\textsc{[Var]}] {\Gamma \vdash
        x:\tau[\vec{\beta}/\vec{\alpha}]}{x:\forall \vec{\alpha}.\tau \in \Gamma \quad \vec{\beta}\; \text{new}} \qquad&
      \infer[\textsc{[Let]}] {\Gamma \vdash \letin{x}{e_1}{e_2}:\tau_2}
            {\Gamma \vdash e_1:\tau_1 \quad \Gamma.x:\forall
              \vec{\alpha}.\tau_1 \vdash e_2:\tau_2 \quad  \vec{\alpha} = \mathit{fv}(\tau_1) \setminus \mathit{fv}(\Gamma)}
    \end{array}
\end{gather*}
\caption{Typing rules for \lang}
\label{fig:typing}
\end{figure*}

\else
\input{typerules.tex}
\fi

\subsection{Minimum Error Source}
\label{sec:minerror}
The objective of this paper is the problem of finding a minimum error source for
a given program $\prog$ subject to a given cost function~\cite{minerrors}.
The problem  formalizes the process of replacing ill typed subexpressions
in a program $\prog$ by $\hole$ to get a well typed program $\prog'$
and associates a cost to each such transformation.

A \emph{location} $\ell$ in a \lang expression $e$ is a path in the
abstract syntax tree of $e$ starting at the root of $e$. The set of all locations of an expression $e$
in a program $p$ is denoted $\locs_p(e)$.  We omit the subscript $p$ when
the program is clear from the context. Each location $\ell$ uniquely identifies a
subexpression $e(\ell)$ within $e$. When an expression $e^{\prime}$ is
clear from the context (typically $e'$ is the whole program $\prog$),
we write $e^\ell$ to denote that $e$ is at a location $\ell$ in
$e^{\prime}$. Similarly, we write $\locs(\ell)$ for $\locs(e'(\ell))$.

The mask function $\mask$ takes an expression $e$ and a location
$\ell \in \locs(e)$ and produces the expression where $e(\ell)$
is replaced by $\hole$ in $e$. 
(Note that $\mask(e,\ell)$ also masks any subexpression of $e(\ell)$.)
We extend $\mask$ to work over an expression $e$
and a set of locations $L \subseteq \locs(e)$.

\begin{definition}[Error source]
  Let $\prog$ be a program.
  A set of locations $L \subseteq \locs(\prog)$ is an \emph{error source} of
  $\prog$ if $\mask(\prog, L)$ is well typed.
\end{definition}

A \emph{cost function} is a mapping $R$ from
a program $\prog$ to a partial function that assigns a positive
weight to locations, $R(p) : \locs(p) \pto \NN^+$.
A location $\ell$ that is not in the domain of $R(\prog)$
is considered to be a \emph{hard} constraint, $\ell \not\in \dom(R(\prog))$.
Hard constraints provide a way for $R$ to specify that a location $\ell$
is not considered to be a source of an error. We require that a location
corresponding to the root node of the program AST cannot be set as hard. 
In other words, for all programs $\prog$ and cost functions $R$ it must be that
$\prog(\ell)=\prog \implies \ell \in \dom(R(\prog))$. This way, we
make sure that there is always at least one error source for an ill-typed
program: the one that masks the whole program. 

Cost functions are extended to a set of locations $L$
in the natural way:
\begin{equation}
  R(\prog)(L) = \sum_{\ell \in L, \ell \in \dom(R(p))} R(p)(\ell)\enspace.
\end{equation}
The minimum error sources are the sets of locations that are error
source and minimize a given cost function.

\begin{definition}[Minimum error source]
  An error source $L \subseteq \locs(\prog)$ for a program $\prog$
  is a \emph{minimum error source} with respect to 
  a cost function $R$ if for any other error source
  $L'$ of $\prog$ $R(p)(L) \leq R(\prog)(L')$.
\end{definition}

In our previous paper~\cite{minerrors}, we used a slightly more
restrictive definition of error source. Namely, we required that an
error source must be minimal, i.e., it does not have a proper subset
that is also an error source. The above definitions imply that a
minimum error source is also minimal since we require that the weights
assigned by cost functions are positive.

\subsection{Satisfiability}
\label{sec:sat}
The classic CNF-SAT problem takes as input a finite set of
propositional clauses $\Clauses$.
A clause is a finite set of literals, which are propositional
variables or negations of propositional variables.
A propositional model $\Assignment$ assigns all propositions into
$\{\True,\False\}$.
An assignment $\Assignment$ is said to satisfy a propositional variable $P$,
written $\Assignment \models P$,
if $\Assignment$ maps $P$ to $\True$.
Similarly, $\Assignment \models \lnot P$ if $\Assignment$ maps $P$
to $\False$.
A clause $C$ is satisfied by $\Assignment$ if at least one literal
in $C$ is satisfied.
The CNF-SAT problem asks if there exists a propositional
model $\Assignment$ that satisfies all clauses in $\Clauses$
simultaneously $\Assignment \models \Clauses$.

\subsection{MaxSAT and Variants}
\label{sec:maxsat}
The MaxSAT problem takes as input a finite set of propositional
soft clauses $\SClauses$ and finds a propositional model $\Assignment$
that maximizes the number of clauses $K$ that are simultaneously
satisfied~\cite{LiM09}.
The \emph{partial MaxSAT} problem
adds a set of \emph{hard} clauses $\HClauses$  that must be satisfied.
The \emph{weighted partial MaxSAT} (WPMaxSAT) problem additionally takes
a map $\Weight$ from soft clauses to positive integer weights and
produces assignments of maximum weight: 

\begin{equation}
  \begin{array}{l}
    \Problem{WPMaxSAT}(\HClauses, \SClauses, \Weight) = \\
    \quad
    \Maximize \sum_{c \in \Clauses} \Weight(c)
    \text{ where }
    \Assignment \models \Clauses \cup \HClauses \text{ and } \Clauses \subseteq \SClauses
  \end{array}\label{eqn:pwmaxsat}
\end{equation}

\subsection{SMT \& MaxSMT}\label{sec:maxsmt}
The \emph{weighted partial MaxSMT} problem (WPMaxSMT)
is formalized
by directly lifting the WPMaxSAT formulation to Satisfiability Modulo Theories
(SMT)~\cite{BSST09}.
The SMT problem takes as input a finite set of assertions $\Assertions$
where each assertion is a first-order formula.
The functions and predicates in the assertions are interpreted
according to a fixed first-order theory $\Th$.
The theory $\Th$ enforces the semantics of the functions to behave
in a certain fashion by restricting the class of first-order
models.
A first-order model $\Assignment$, in addition to assigning
variables to values in a domain, assigns semantics to
the function symbols over the domain.
As an example, the theory of linear real arithmetic enforces
the domain to be the mathematical real numbers $\mathcal{R}$ and the
built-in function symbol $+$ to behave as the mathematical
plus function.
The model $\Assignment$ is said to satisfy a formula $\phi$,
written again as $\Assignment \models \phi$,
if $\phi$ evaluates to $\True$ in $\Assignment$.
We consider a theory $\Th$ to be a class
of models.
A formula (or finite set of formulas) is satisfiable
modulo $\Th$, written as $\Assignment \modelsTh \phi$,
if there is a model $\Assignment$ such that
$\Assignment \in \Th$ and $\Assignment \models \phi$.\footnote{
  This informal introduction ignores many aspects of SMT
  such as non-standard models for the theory of reals.
}

Most concepts directly generalize from MaxSAT to MaxSMT:
satisfiability is now modulo the models of $\Th$, and
soft and hard clauses are now over $\Th$-literals.
Many SMT solvers are organized around adding $\Th$-valid
formulas, known as theory lemmas, into $\Lemmas$ to refine the search.
(Thus $\Lemmas$ still only contains formulas entailed by $\Assertions$.)
The optimization formulation of WPMaxSMT is nearly identical
to~\eqref{eqn:pwmaxsat}:

\begin{equation}
  \begin{array}{l}
    \Problem{WPMaxSMT}(\HAssertions, \SAssertions, \Weight) = \\
    \quad
    \Maximize \sum_{c \in \Assertions} \Weight(c)
    \text{ where }
    \Assignment \modelsTh \Assertions \cup \HAssertions \text{ and } \Assertions \subseteq \SAssertions
  \end{array}\label{eqn:pwmaxsmt}
\end{equation}

We reduce computing minimum error sources to solving WPMaxSMT problems.
We first generate typing constraints from the given input program that are
satisfiable iff the input program is well typed.
We then specify the weight function $w$ by labeling a subset of the
assertions according to the cost function $R$. 

\subsection{Theory of Inductive Datatypes}\label{sec:idt}
The theory of inductive data types~\cite{idt} allows us to compactly express the
needed typing constraints.
The theory allows for users to define their own inductive data types and
state equality constraints over the terms of that data type.
We define an inductive data type $\mathsf{Types}$ that represents the ground monotypes of \lang:
\begin{align}
  \qquad t \in \mathsf{Types} ::= \tint \mid \tbool \mid \fun{t}{t}
  \label{eqn:types}
\end{align}
Here, the term constructor $\funconstr$ is used to encode the
ground function types.
The models of the theory of inductive data types forces the interpretation of the constructors
in the expected fashion.
For instance:
\begin{enumerate}
\item Different constructors produce disequal terms.
  \[ \tint \neq \tbool, \forall \alpha,\beta. \tbool \neq \fun{\alpha}{\beta} \land \tint \neq \fun{\alpha}{\beta} \]
\item Every term is constructed by some constructor.
  \[ t = \tbool \lor t = \tint \lor \exists \alpha,\beta. t = \fun{\alpha}{\beta} \]
\item The constructors are injective.
  \[ \forall \alpha, \beta, \gamma, \delta \in \mathsf{Types}.\;
  \fun{\alpha}{\beta}\!=\!\fun{\gamma}{\delta} 
  \Rightarrow \alpha\!=\!\gamma \wedge \beta\!=\!\delta \]
\end{enumerate}
Thus, the theory enforces that the ground monotypes of \lang
are faithfully interpreted by the terms of $\mathsf{Type}$.

To support typing expressions such as $(a, b, \_)$ and others found in
realistic languages, we extend $\mathsf{Types}$ in \eqref{eqn:types}
with additional type constructors, e.g., $\mathsf{product}(t,t)$, to
encode product types $\tau_1 * \tau_2$ and user-defined algebraic data
types.  This pre-processing pass is straightforward but outside of the
scope of this paper.

\section{Algorithm}
\label{sec:algorithm}
We now introduce a refinement of the typing relation used in~\cite{minerrors}
to generate typing constraints.
The novelty of this new typing relation is the ability to specify
a set of variable usage locations whose typing constraints
are abstracted as the principal type of the variable. We then
describe an algorithm that iteratively uses this typing relation
to find a minimum error source while expanding only those
principal type usages that are relevant for the minimum source.

\subsection{Notation and Setup}
\label{sec:principal}
Standard type inference implementations handle
expressions of the form
$\letin{x}{e_1}{e_2}$
by computing the \emph{principal} type of $e_1$,
binding $x$ to the principal type $\sigma_p$ in
the environment $\Gamma.x:\sigma_p$,
and proceeding to perform type inference on $e_1$~\cite{principal}.
Given an environment $\Gamma$,
the type $\sigma_p$ is the principal type for
$e$ if $\Gamma \vdash e : \sigma_p$ and for any other $\sigma$ such that $\Gamma \vdash e : \sigma$ then
$\sigma$ is a generic instance of $\sigma_p$.
Note that a principal
type is unique, subject to $e$ and $\Gamma$, up to the renaming of bound
type variables in $\sigma_p$.

We now introduce several auxiliary functions and sets that we use in our algorithm. We 
define $\Principal$ to be a partial function accepting an expression $e$ and a typing 
environment $\Gamma$ where $\Principal(\Gamma, e)$ returns a principal type of $e$ subject to $\Gamma$.
If $e$ is not typeable in $\Gamma$, then $(\Gamma, e) \not \in \dom(\Principal)$. 
Next, we define a mapping $\Ulocs$ for the usage locations of a variable.
Formally, $\Ulocs$ is a partial function such that given a location $\ell$
of a \texttt{let} variable definition and a program $p$ returns the set $\Ulocs_{p}(\ell)$
of all locations where this variable is used in $p$. Note that a location of a \texttt{let}
variable definition is a location corresponding to the root of the defining expression.
We also make use of a function for the definition location $\Dlocs$.
The function $\Dlocs$ reverses the mapping of $\Ulocs$ for a variable usage.
More precisely, $\Dlocs(p, \ell)$ returns the location where the variable
appearing at $\ell$ was defined in $p$. Also, for a set of locations $L$
we define $\Vlocs(\ell)$ to be the set of all locations in
$\locs(\ell)$ that correspond to usages of \texttt{let} variables.

For the rest of this section, we assume a fixed program $p$ for which
the above functions and sets are precomputed. We do not provide
detailed algorithms for computing these functions since they are
either straightforward or well-known from the literature. For instance, the
$\Principal$ function can be implemented using the classical W
algorithm~\cite{principal}.

\subsection{Constraint Generation}
\label{sec:constraint}

\ifARXIV
\begin{figure*}[p]
  \centering
  \begin{tabular}{c}
  \infer[\textsc{[A-Abs]}] { \Pi, \Gamma \trel{\explocs} (\lambda
    x.e)^{\ell}: \gamma \mid \{\prop_{\ell} \Rightarrow (\{
    \gamma=\fun{\alpha}{\beta}\} \cup \Assertions)\}
  }{
    \Pi.x:\alpha,\Gamma.x:\alpha \trel{\explocs} e: \beta \mid \Assertions &\quad
    \gamma\; \text{new}
  }\\[0.8cm]
  
  \infer[\textsc{[A-App]}] { \Pi,\Gamma \trel{\explocs}  (e_1 \; e_2)^{\ell}:
    \gamma \mid \{\prop_{\ell} \Rightarrow
    (\{\alpha=\fun{\beta}{\gamma}\} \cup \Assertions_1 \cup
    \Assertions_2)\}
  }{
    \Pi,\Gamma \trel{\explocs} e_1: \alpha \mid \Assertions_1 &\quad
    \Pi,\Gamma \trel{\explocs} e_2: \beta \mid \Assertions_2 &\quad
    \gamma\; \text{new}
  }\\[0.8cm]

  \infer[\textsc{[A-Cond]}] {
    \Pi,\Gamma \trel{\explocs}
    (\ite{e_1^{\ell_1}}{e_2^{\ell_2}}{e_3^{\ell_3}})^\ell: \gamma \mid
    \{\prop_\ell \Rightarrow
    (\Assertions_1 \cup \Assertions_2 \cup \Assertions_3 \cup \Assertions_4)\}
  }{
    \def\arraystretch{1.5}
    \begin{array}{l}
      \Pi,\Gamma \trel{\explocs} e_1: \alpha_1 \mid \Assertions_1 \qquad
      \Pi,\Gamma \trel{\explocs} e_2: \alpha_2 \mid \Assertions_2 \qquad
      \Pi,\Gamma \trel{\explocs} e_3: \alpha_3 \mid \Assertions_3 \qquad
      \gamma\; \text{new} \\
      \Assertions_4 =
      \{(\prop_{\ell_1} \Rightarrow \alpha_1=\tbool),
      (\prop_{\ell_2} \Rightarrow \alpha_2=\gamma), 
      (\prop_{\ell_3} \Rightarrow \alpha_3=\gamma)\}
    \end{array}
  } \\[0.8cm]

  \infer[\textsc{[A-Hole]}] {
    \Pi,\Gamma \trel{\explocs} \hole: \alpha \mid \emptyset
  }{
    \alpha\; \text{new}
  } \\[0.8cm]

  \infer[\textsc{[A-Bool]}] {
    \Pi,\Gamma \trel{\explocs} b^\ell: \alpha \mid \{\prop_\ell \Rightarrow \alpha=\tbool\}
  }{
    b \in \mathbb{B} &\quad \alpha\; \text{new}
  }\\[0.8cm]
  
  \infer[\textsc{[A-Int]}]  {
    \Pi,\Gamma \trel{\explocs} n^\ell: \alpha \mid \{\prop_\ell \Rightarrow \alpha=\tint\}
  }{
    n \in \mathbb{Z} &\quad \alpha\; \text{new}
  } \\ [0.8cm]
  
  \infer[\textsc{[A-Var-Exp]}] {\Pi,\Gamma \trel{\explocs} x^\ell:
    \gamma \mid \{\prop_{\ell} \Rightarrow (\{
    \gamma=\alpha[\vec{\beta}/\vec{\alpha}]\} \cup
    \Assertions[\vec{\beta}/\vec{\alpha}])\}}
  { \ell \in \explocs &\quad
    x:\forall \vec{\alpha}.(\Assertions \Rrightarrow \alpha) \in \Gamma &\quad
    \vec{\beta}, \gamma\; \text{new}
  }\\[0.8cm]

  \infer[\textsc{[A-Var-Prin]}]{
    \Pi,\Gamma \trel{\explocs} x^\ell:
    \gamma \mid \{\prop_{\ell} \Rightarrow (\{
    \gamma=\alpha[\vec{\beta}/\vec{\alpha}]\} \cup
    \Assertions[\vec{\beta}/\vec{\alpha}])\}
  }{
    \ell \not \in \explocs &\quad
    x:\forall \vec{\alpha}.(\Assertions \Rrightarrow \alpha) \in \Pi &\quad
    \vec{\beta}, \gamma\; \text{new}
  }\\[0.8cm]

  \infer[\textsc{[A-Let-Exp]}] {
    \Pi,\Gamma \trel{\explocs} (\letin{x}{e_1^{\ell_1}}{e_2})^{\ell}: \gamma \mid \{\prop_\ell \Rightarrow (\{\gamma = \alpha_2\} \cup \Assertions_1[\vec{\beta}/\vec{\alpha}] \cup \Assertions_2)\}
  }{
    \def\arraystretch{1.5}
    \begin{array}{l}
      \ell_1 \in \explocs
      \\
      \Pi, \Gamma \trel{\explocs} e_1: \alpha_1 \mid \Assertions_1
      \qquad
      \vec{\alpha} = \fv(\Assertions_1) \setminus \fv(\Gamma)
      \qquad
      \tau_{exp} = \forall \vec{\alpha}.(\Assertions_1 \Rrightarrow \alpha_1)
      \\
      \Pi,\Gamma.x: \tau_{exp}
      \trel{\explocs} e_2: \alpha_2 \mid \Assertions_2
      \qquad
      \vec{\beta},\gamma \text{ new}
    \end{array}
  }\\[0.8cm]


  \infer[\textsc{[A-Let-Prin]}] {
    \Pi,\Gamma \trel{\explocs} (\letin{x}{e_1^{\ell_1}}{e_2})^{\ell}:
    \gamma \mid \{\prop_\ell \Rightarrow (\{\gamma = \alpha_2\} \cup \Assertions_1[\vec{\beta}/\vec{\alpha}] \cup \Assertions_2)
    \}
  }{
    \def\arraystretch{1.5}
    \begin{array}{l}
      \ell_1 \not\in \explocs
      \\
      \Principal(\Pi, e_1) = \forall \vec{\delta}. \tau_{p}
      \qquad
      \alpha \text{ new}
      \qquad
      \tau_{prin} = \forall \alpha, \vec{\delta}.(\{\UsePrincipal{\ell_1} \Rightarrow
      \alpha = \tau_{p}\} \Rrightarrow \alpha)
      \\
      \Pi, \Gamma \trel{\explocs} e_1: \alpha_1 \mid \Assertions_1
      \qquad
      \vec{\alpha} = \fv(\Assertions_1) \setminus \fv(\Gamma)
      \qquad
      \tau_{exp} = \forall \vec{\alpha}.(\Assertions_1 \Rrightarrow \alpha_1)
      \\
      \Pi.x: \tau_{prin},\, \Gamma.x: \tau_{exp}
      \trel{\explocs} e_2: \alpha_2 \mid \Assertions_2 \qquad
      \vec{\beta}, \gamma \text{ new}
    \end{array}
  }
\end{tabular}
\caption{Rules defining the constraint typing relation for \lang}
\label{fig:consgen}
\end{figure*}

\else
\input{constraintgen.tex}
\fi

The main idea behind our algorithm, described in
Section~\ref{sec:iter-alg}, is to iteratively discover which principal
type usages must be expanded to compute a minimum error source. The
technical core of the algorithm is a new typing relation that produces
typing constraints subject to a set of locations where principal type
usages must be expanded.

We use $\Assertions$ to denote a set of logical assertions in the signature 
of $\mathsf{Types}$ that represent typing constraints.  Henceforth, when we 
refer to types we mean terms over $\mathsf{Types}$. \emph{Expanded locations} 
are a set of locations $\explocs$ such that $\explocs \subseteq \locs(\prog)$.  
Intuitively, this is a set of locations corresponding to usages of 
$\Let$ variables $x$ where the typing of $x$ in the current iteration
of the algorithm is expanded into the corresponding typing constraints.
Those  locations of usages of $x$ that are not expanded will treat $x$ using its 
principal type.
We also introduce a set of locations whose usages must be expanded $\explocs_0$.
We will always assume $\explocs_0 \subseteq \explocs$.
Formally, $\explocs_0$ is the set of all  program locations in $p$ except the locations
of well-typed \texttt{let} variables and their usages.
This definition enforces that usages of variables that have no principal type are always expanded.
In summary, $\explocs_0 \subseteq \explocs \subseteq \locs(\prog)$.

We define a typing relation $\trel{\explocs}$ over
$(\Pi ,\Gamma, e, \alpha, \Assertions)$ which is parameterized by
$\explocs$. The relation is given by judgments of the form:
\[ \Pi, \Gamma \trel{\explocs} e : \alpha \mid \Assertions.
\] 
Intuitively, the relation holds iff expression $e$ in $\prog$ has type
$\alpha$ under typing environment $\Gamma$ if we solve the constraints
$\Assertions$ for $\alpha$. (We make this statement formally precise
later.) The relation depends on $\explocs$, which controls
whether a usage of a $\Let$ variable is typed by the principal
type of the $\Let$ definition or the expanded typing constraints of
that definition.

For technical reasons, the principal types are computed in tandem
with the expanded typing constraints. This is because both the
expanded constraints and the principal types may refer to type
variables that are bound in the environment, and we have to ensure that
both agree on these variables.
We therefore keep track of two separate typing environments:
\begin{itemize}
\item the environment $\Pi$ binds $\Let$
variables to the principal types of their defining expressions if the
principal type exists with respect to $\Pi$, and
\item the typing environment $\Gamma$ binds
$\Let$ variables to their expanded typing constraints (modulo
$\explocs$).
\end{itemize}
The typing relation ensures that the two environments are
kept synchronized. To properly handle polymorphism, the bindings in
$\Gamma$ are represented by typing schemas:
\[x: \forall \vec{\alpha}.(\Assertions \Rrightarrow \alpha)
\] 
The schema states that $x$ has type $\alpha$ if we solve the typing
constraints $\Assertions$ for the variables $\vec{\alpha}$. To simplify
the presentation, we also represent bindings in $\Pi$ as type schemas.
Note that we can represent an arbitrary type $t$ by the schema
$\forall \alpha. (\{\alpha = t\} \Rrightarrow \alpha)$ where $\alpha
\notin \fv(t)$.
The symbol $\Rrightarrow$ is used here to suggest,
but keep syntactically separate, the notion of logical implication
$\Rightarrow$ that is implicitly present in the schema.

The typing relation $\Pi,\Gamma \trel{\explocs} e: \alpha \mid \Assertions$ is
defined in Figure~\ref{fig:consgen}.
It can be seen as a constraint generation procedure that goes over an
expression $e$ at location $\ell$ and generates a set of typing
constraints $\Assertions$.  For the purpose of computing error
sources, we associate with each location $\ell$ a propositional
variable $\prop_\ell$. The location $\ell$ is in the computed error
source iff the variable $\prop_\ell$ is assigned to $\False$. This is
also reflected in the typing constraints. All typing constraints added
at location $\ell$ are guarded by the variable $\prop_\ell$. That is,
the clauses $\assertion_n$ in the constraint generated for an
expression $e^{\ell_n}$ with a subexpression at location $\ell_1$ have
the rough form:
\[ \prop_{\ell_n} \Rightarrow \cdots \Rightarrow \prop_{\ell_{1}} \Rightarrow
  \alpha_1 = t
\]
where $\alpha_1 = t$ is the typing constraint on the subexpression
$\ell_1$.  The $\prop_{\ell_i}$ are the propositional variables
associated with the locations on the path from $\ell_n$ to $\ell_1$ in
the abstract syntax tree. Only if $\prop_{\ell_n}, \ldots,
\prop_{\ell_1}$ are all true, is the constraint $\alpha_1 = t$
active.  If any of the variables $\prop_{\ell_i}$ is false,
$\assertion_n$ is trivially satisfied.  This captures the fact that
the typing constraint of the subexpression at $\ell_1$ should be
disregarded if any of the expressions $e^{\ell_i}$ in which it is
contained are part of the error source (i.e., $e^{\ell_i}$
is replaced by a hole expression, and with it $e_1$).

The rules \textsc{A-Let-Prin} and \textsc{A-Let-Exp} govern the
computation and binding of typing constraints and principal types for
$\Let$ definitions $(\letin{x}{e_1^{\ell_1}}{e_2})^\ell$. If $e_1$ has
no principal type under the current environment $\Pi$, then $\ell_1
\in \explocs$ by the assumption that $\explocs_0 \subseteq
\explocs$. Thus, when rule \textsc{A-Let-Prin} applies,
$\Principal(\Pi,e_1)$ is defined. The rule then binds $x$ in $\Pi$ to
the principal type and binds $x$ in $\Gamma$ to the expanded typing
constraints obtained from $e_1$. 

The \textsc{[A-Let-Prin]} rule binds $x$ in both $\Pi$ and $\Gamma$ as it is possible that
in the current iteration some usages of $x$ need to be typed with
principal types and some with expanded constraints. For instance, our
algorithm can expand usages of a function, say $f$, in the first
iteration, and then expand all usages of, say $g$, in the next
iteration. If $g$'s defining expression in turn contains calls to $f$,
those calls will be typed with principal types. This is done because
there may exist a minimum error source that does not require that the
calls to $f$ in $g$ are expanded.

After extending the typing environments, the rule recurses to compute
the typing constraints for the body $e_2$ with the extended
environments. Note that the rule introduces an auxiliary propositional
variable $\UsePrincipal{\ell_1}$ that guards all the typing
constraints of the principal type before $x$ is bound in $\Pi$. This
step is crucial for the correctness of the algorithm. We refer to the
variables as \emph{principal type correctness
  variables}. That is, if $\UsePrincipal{\ell_1}$ is $\True$ then this
means that the definition of the variable bound at $\ell_1$ is not
involved in the minimum error source and the principal type 
safely abstracts the associated unexpanded typing constraints.


The rule \textsc{A-Let-Exp} applies whenever $\ell_1 \in \explocs$.
The rule is almost identical to the \textsc{A-Let-Prin} rule, except
that it does not bind $x$ in $\Pi$ to $\tau_{prin}$ (the principal type).
This will have the effect that for all usages of $x$ in $e_2$,
the typing constraints for $e_1$ to which $x$ is bound in $\Gamma$
will always be instantiated.
By the way the algorithm extends the set $\explocs$, $\ell_1 \in
\explocs$ implies that $\ell_1 \in \explocs_0$, i.e., the defining
expression of $x$ is ill-typed and does not have a principal type.

The \textsc{A-Var-Prin} rule instantiates the typing constraints of
the principal type of a $\Let$ variable $x$ if $x$ is bound in
$\Pi$ and the location of $x$ is not marked to be
expanded. Instantiation is done by substituting the type variables
$\vec{\alpha}$ that are bound in the schema of the principle type with
fresh type variables $\vec{\beta}$. The \textsc{A-Var-Exp} rule is
again similar, except that it handles all usages of $\Let$ 
variables that are marked for expansion, as well as all usages of
variables that are bound in lambda abstractions.

The remaining rules are relatively straightforward. The rule
\textsc{A-Abs} is noteworthy as it simultaneously binds the
abstracted variable $x$ to the same type variable $\alpha$ in both
typing environments. This ensures that the two environments
consistently refer to the same bound type variables when they are used
in the subsequent constraint generation and principal type computation
within $e$.

\subsection{Reduction to Weighted Partial MaxSMT}
\label{sec:reduction-wmaxsmt}

Given a cost function $R$ for program $p$ and a set of locations
$\explocs$ where $\explocs_0 \subseteq \explocs$, we generate a
WPMaxSMT instance $\instance(p,R,\explocs) = (\HAssertions,
\SAssertions, \Weight)$ as follows.  Let $\Assertions_{p, \explocs}$
be a set of constraints such that $\emptyset, \emptyset
\trel{\explocs} p: \alpha \mid \Assertions_{p, \explocs}$ for some
type variable $\alpha$. Then define
\begin{align*}
  \HAssertions =\; & \Assertions_{p, \explocs} 
  \cup \pset{\prop_\ell}{\ell \notin \dom(R(p))} 
  \cup  \pdefs(p)
     \\
  \SAssertions =\; & \pset{\prop_\ell}{\ell \in \dom(R(p))}\\
  w(\prop_\ell) =\; & R(p)(\ell), \; \text{for all $\prop_\ell \in \SAssertions$}
\end{align*}
The set of assertions $\pdefs(p)$ contains the definitions
for the principal type correctness variables $P_\ell$. For a
\texttt{let} variable $x$ that is defined at some location
$\ell$, the variable $\UsePrincipal{\ell}$ is defined to be $\True$
iff
\begin{itemize}
\item each location variable $\prop_{\ell'}$ for a location $\ell'$
  in the defining expression of $x$ is $\True$, and
\item each principal type correctness variable $\UsePrincipal{\ell'}$
  for a \texttt{let} variable that is defined at $\ell'$ and
  used in the defining expression of $x$ is $\True$.
\end{itemize}
Formally, $\pdefs(p)$ defines the set of formulas
\begin{align*}
   \pdefs(p) = \; & \pset{\pdef_{\ell}}{\ell \in
     \dom(\Ulocs_p)}\\
   \pdef_{\ell} = \; & \left(
   \UsePrincipal{\ell} \Leftrightarrow \bigwedge_{\ell' \in \locs(\ell)} \prop_{\ell'} \;\;\;\wedge 
   \bigwedge_{\ell' \in \Vlocs(\ell)}
   \UsePrincipal{\Dlocs(\ell')} \right)
\end{align*}
Setting the $P_\ell$ to $\False$ thus captures all possible
error sources that involve some of the locations in the defining
expression of $x$, respectively, the defining expressions of other
variables that $x$ depends on.
Recall that the propositional variable $\UsePrincipal{\ell}$ is used
to guard all the instances of the principal types of $x$ in
$\Assertions_{p, \explocs}$. Thus, setting $\UsePrincipal{\ell}$ to
$\False$ will make all usage locations of $x$ well-typed that have not
yet been expanded and are thus constrained by the principal
type. By the way $\UsePrincipal{\ell}$ is defined, the cost of setting
$\UsePrincipal{\ell}$ to $\False$ will be the minimum weight of all
the location variables for the locations of $x$'s
definition and its dependencies. Thus, $\UsePrincipal{\ell}$
conservatively approximates all the potential minimum error sources
that involve these locations.

We denote by $\minerrors$ the procedure that given $p$, $R$, and
$\explocs$  returns some model $\Assignment$ that is a
solution of $\instance(p,R,\explocs)$.

\begin{lemma}
  \label{lem:minerrors-total}
  $\minerrors$ is total.
\end{lemma}

Lemma~\ref{lem:minerrors-total} follows from our
assumption that $R$ is defined for the root location $\ell_p$ of
the program $p$. That is, $\instance(p,R,\explocs)$ always has some
solution since $\HAssertions$ holds in any model $\Assignment$
where $\Assignment \not \models \prop_{\ell_p}$.

Given a model $\Assignment = \minerrors(p, R, \explocs)$,
we define $L_{\Assignment}$ to be the set of locations excluded in
$\Assignment$:
\[L_{\Assignment} = \pset{\ell \in \locs(p)}{\Assignment \models \neg \prop_\ell}.
\] 

\subsection{Iterative Algorithm}
\label{sec:iter-alg}

Next, we present our iterative algorithm for computing minimum type
error sources. 

In order to formalize the termination condition of the algorithm, we
first need to define the set of usage locations of \texttt{let} 
variables in program $p$ that are in the scope of the current
expansion $\explocs$. We denote this set by $\scope(p,
\explocs)$. Intuitively, $\scope(p, \explocs)$ consists of all those
usage locations of \texttt{let} variables that either occur in the
body of a top-level \texttt{let} declaration or in the defining
expression of some other \texttt{let} variable which has at
least one expanded usage location in $\explocs$.
Formally, $\scope(p, \explocs)$ is the largest set of usage locations
in $p$ that satisfies the following condition: for all $\ell \in \dom(\Ulocs_p)$, 
if $\Ulocs_p(\ell) \cap \explocs = \emptyset \wedge \Ulocs_p(\ell) \not = \emptyset$, 
then $\locs(\ell) \cap \scope(p, \explocs) = \emptyset$.

For $\Assignment = \minerrors(p, R, \explocs)$, we then define $\usages(p,\explocs,\Assignment)$ 
to be the set of all usage locations of the
$\Let$ variables in $p$ that are in scope of the current expansions and
that are marked for expansion. That is, $\ell \in \usages(p,\explocs,M)$
iff
\begin{enumerate}
\item $\ell \in \scope(p, \explocs)$, and
\item $\Assignment \not \models \UsePrincipal{\Dlocs(\ell)}$
\end{enumerate}
Note that if the second condition holds, then a
potentially cheaper error source exists that involves locations in the
definition of the variable $x$ used at $\ell$. Hence, that usage of $x$
should not be typed by $x$'s principal type but by the expanded typing
constraints generated from $x$'s defining expression.

We say that a solution $L_M$, corresponding to the result of $\minerrors(p,R,\explocs)$, is
\emph{proper} if $\usages(p,\explocs,M) \subseteq \explocs$, i.e., $L_M$
does not contain any usage locations of $\Let$ variables that
are in scope and still typed by unexpanded instances of principal types.

Algorithm~\ref{alg:algorithm} shows the iterative algorithm. It takes
an ill-typed program $\prog$ and a cost function $R$ as input and
returns a minimum error source.  The set $\explocs$ of locations to be
expanded is initialized to $\explocs_0$. In each iteration, the
algorithm first computes a minimum error source for the current
expansion using the procedure $\minerrors$ from the previous
section. If the computed error source is proper, the algorithm
terminates and returns the current solution $L_M$. Otherwise, all usage
locations of $\Let$ variables involved in the current minimum
solution are marked for expansion and the algorithm continues.

\begin{algorithm}[t]
\begin{algorithmic}[1]
\Procedure{IterMinError}{$p,R$}
\State $\explocs \gets \explocs_0$
\Loop
\State $M \gets \minerrors(p, R,\explocs)$
\State $L_u \gets \usages(p, \explocs, M)$
\If{$L_u \subseteq \explocs$} \label{alg:termination}
\State \textbf{return} $L_M$ 
\EndIf
\State $\explocs \gets \explocs \cup L_u$
\EndLoop
\EndProcedure
\end{algorithmic}
\caption{Iterative algorithm for computing a minimum error source}\label{alg:iter-min-errs}
\label{alg:algorithm}
\end{algorithm}

\subsection{Correctness}

We devote this section to proving the correctness of our iterative
algorithm. In a nutshell, we show by induction  that the solutions computed
by our algorithm are also solutions of the naive algorithm that
expands all usages of \texttt{let} variables immediately as
in~\cite{minerrors}.

We start with the base case of the induction where we fully expand all
constraints, i.e., $\explocs=\locs(p)$.

\begin{lemma}
  \label{thm:expanded-correctness}
  Let $p$ be a program and $R$ a cost function and let $M =
  \minerrors(p,R,\locs(p))$. Then $L_M \subseteq \locs(p)$ is a
  minimum error source of $p$ subject to $R$.
\end{lemma} 

Lemma~\ref{thm:expanded-correctness} follows from
\cite[Theorem~1]{minerrors} because if $\explocs=\locs(p)$, then we
obtain exactly the same reduction to WPMaxSMT as in our previous
work. More precisely, in this case the \textsc{A-Var-Prin} rule is
never used. Hence, all usages of \texttt{let} variables are
typed by the expanded typing constraints according to rule
\textsc{A-Var-Exp}.  The actual proof requires a simple induction over
the derivations of the constraint typing relation defined in
Figure~\ref{fig:consgen}, respectively, the constraint typing relation
defined in \cite[Figure 4]{minerrors}.

We next prove that in order to achieve full expansion it is not
necessary that $\explocs = \locs(p)$. To this end, define the set
$\explocs_p$, which consists of $\explocs_0$ and all usage locations
of $\Let$ variables in $p$:
\[\explocs_p = \explocs_0 \;\cup\; \bigcup_{l \in \dom(\Ulocs_{p})} \Ulocs_{p}(l).\]
Then $\vdash_\explocs$ generates the same constraints as
$\vdash_{\locs(p)}$ as stated by the following lemma.

\begin{lemma}
  \label{lemma:expanded-correctness-relaxed}
  For any $p$, $\Pi$, $\Gamma$, $\alpha$, and $\Assertions$, we have
  $\Pi, \Gamma \vdash_{\explocs_p} p:\alpha \;|\; \Assertions$ iff
  $\;\Pi, \Gamma \vdash_{\locs(p)} p:\alpha \;|\; \Assertions$.
\end{lemma}

Lemma~\ref{lemma:expanded-correctness-relaxed} can be proved using a
simple induction on the derivations of $\vdash_{\explocs_p}$,
respectively, $\vdash_{\locs(p)}$. First, note that $\locs(p)
\setminus \explocs_p$ is the set of locations of well-typed
\texttt{let} variable definitions in $p$. Hence, the derivations using
$\vdash_{\explocs_p}$ will never use the \textsc{A-Let-Exp} rule, only
\textsc{A-Let-Prin}. However, the \textsc{A-Let-Prin} rule updates
both $\Pi$ and $\Gamma$, so applications of \textsc{A-Var-Exp}
(\textsc{A-Var-Prin} is never used in either case) will be the same as
if $\vdash_{\locs(p)}$ is used.

The following lemma states that if the iterative algorithm terminates,
then it computes a correct result.

\begin{lemma}
  \label{thm:correctness}
  Let $p$ be a program, $R$ a cost function, and $\explocs$ such that
  $\explocs_0 \subseteq \explocs \subseteq \explocs_p$. Further, let
  $M = \minerrors(p,R,\explocs)$ such that $L_M$ is proper. Then, 
  $L_M$ is a minimum error source of $p$ subject to $R$.
\end{lemma}

The proof of Lemma~\ref{thm:correctness} can be found in the extended
version of the paper~\cite{minerrors-t}. For brevity, we provide here only the
high-level argument. The basic idea is to show that adding each of the
remaining usage locations to $\explocs$ results in typing constraints
for which $L_{\Assignment}$ is again a proper minimum error source. More precisely,
we show that for each set $\diffset$ such that $\explocs_0 \subseteq
\explocs \subseteq \explocs \cup \diffset \subseteq \explocs_p$, if
$\Assignment$ is the maximum model of $\instance(p, R, \explocs)$
from which $L_M$ was computed, then $\Assignment$ can be extended to a
maximum model $\Assignment'$ of $\instance(p, R, \explocs \cup
\diffset)$ such that $L_{\Assignment'} = L_{\Assignment}$. That is, $L_{\Assignment}$ is again a
proper minimum error source for $\instance(p, R, \explocs \cup
\diffset)$.  The proof goes by induction on the cardinality of the set
$\diffset$. Therefore, by the case
$\explocs \cup \diffset = \explocs_p$,
Lemma~\ref{thm:expanded-correctness}, and
Lemma~\ref{lemma:expanded-correctness-relaxed} we have that $L_{\Assignment}$ is a
true minimum error source for $p$ subject to $R$.
  
Finally, note that the iterative algorithm always terminates since
$\explocs$ is bounded from above by the finite set $\explocs_p$ and
$\explocs$ grows in each iteration. Together with
Lemma~\ref{thm:correctness}, this proves the total correctness
of the algorithm.

\begin{theorem}
  \label{thm:algo-correctness}
  Let $p$ be a program and $R$ a cost function. Then,
  $\textsc{IterMinError}(p,R)$ terminates and computes a minimum error
  source for $p$ subject to $R$.
\end{theorem}

\section{Implementation and Evaluation}
\label{sec:evaluation}

In this section we describe the implementation of our algorithm that targets
the Caml subset of the OCaml language. We also present the results of evaluating
our implementation on the OCaml student benchmark suite
from~\cite{seminal} and the
GRASShopper~\cite{grasshopper} program verification tool.

The prototype implementation of our algorithm was uniformly faster than the naive approach
in our experiments.
Most importantly, the number of generated typing constraints produced by our algorithm
is almost an order of magnitude smaller than when using the naive approach. Consequently,
our algorithm also ran faster in the experiments. 

We note that the new algorithm and the algorithm in~\cite{minerrors} provide the same 
formal guarantees. Since we made experiments on the quality of type error sources in~\cite{minerrors},
we feel a new evaluation--over largely the same set of benchmarks and the same ranking criterion--would 
not be a significant contribution beyond the work done in~\cite{minerrors}. We refer the reader
to that paper for more details. 
 
\subsection{Implementation}

Our implementation bundles together the EasyOCaml~\cite{ecaml} tool and the MaxSMT 
solver $\nu$Z~\cite{nuZ, newZ}.
The $\nu$Z solver is available as a branch of the SMT solver Z3~\cite{z3}.
We use EasyOCaml for generating typing constraints for OCaml programs.
Once we convert the constraints to the weighted  MaxSMT instances, we use Z3's
weighted MaxRes~\cite{maxres} algorithm to compute a minimum error source.

\paragraph{Constraint Generation.} 
EasyOCaml is a tool that helps programmers debug type errors by
computing a slice of a program involved in the type
error~\cite{horder}. The slicing algorithm that EasyOCaml implements
relies on typing constraint generation. More precisely, EasyOCaml
produces typing constraints for the Caml part of the OCaml language,
including algebraic data types, reference, etc. The implementation of
our algorithm modifies EasyOCaml so that it stores a map
from locations to the corresponding generated typing constraints. This map
is then used to compute the principal types for \texttt{let} variables.
Rather than using the algorithm W, we take typing constraints of
locations within the \texttt{let} defining expression and compute a
most general solution to the constraints using a unification
algorithm~\cite{robinson,pierce}. In other words, principal types for
\texttt{let} defining variables are computed in isolation, with no
assumptions on the bound variables, which are left intact. Then, we
assign each program location with a weight using a fixed cost
function.  The implementation uses a modified version of the cost
function introduced in Section~\ref{sec:overview} where each
expression is assigned a weight equal to its AST size.  The
implemented function additionally annotates locations that come from
expressions in external libraries and user-provided type annotations
as hard constraints.  This means that they are not considered as a
source of type errors.

The generation of typing constraints for each iteration in our
algorithm directly follows the typing rules in
Figure~\ref{fig:consgen}. In addition, we perform a simple
optimization that reduces the total number of typing constraints.
When typing an expression $\letin{x}{e_1}{e_2}$, the
\texttt{A-Let-Prin} and \texttt{A-Let-Exp} rules always add a fresh
instance of the constraint $\Assertions_1$ for $e_1$ to the whole set
of constraints.  This is to ensure that type errors in $e_1$ are not
missed if $x$ is never used in $e_2$.  We can avoid this duplication
of $\Assertions_1$ in certain cases.  If a principal type was
successfully computed for the \texttt{let} variable beforehand, the
constraints $\Assertions_1$ must be consistent.  If the expression
$e_1$ refers to variables in the environment that have been bound by
lambda abstraction, then not instantiating $\Assertions_1$ at all
could make the types of these variables under-constrained. However, if
$\Assertions_1$ is consistent and $e_1$ does not contain variables
that come from lambda abstractions, then we do not need to include a
fresh instance of $\Assertions_1$ in \texttt{A-Let-Prin}. Similarly, if
$e_1$ has no principal type because of a type error and the variable $x$ is
used somewhere in $e_2$, then the algorithm ensures that all such usages are
expanded and included in the whole set of typing constraints.
Therefore, we can safely omit the extra instance of $\Assertions_1$ in
this case as well.


\paragraph{Solving the Weighted MaxSMT Instances.}
Once our algorithm generates typing constraints for an iteration,
we encode the constraints in an extension of the
\texttt{SMT-LIB}~2 language~\cite{smtlib}.
This extension allows us to handle the theory of inductive data types which
we use to encode types and type variables,  whereas locations are encoded as
propositional variables.
We compute the weighted partial MaxSMT solution for the encoded typing constraints by
using Z3's weighted partial MaxSMT facilities.
In particular, we configure the solver to use the MaxRes~\cite{maxres} algorithm
for solving the weighed partial MaxSMT problem.

\subsection{Evaluation}
We evaluated our implementation on the student OCaml benchmarks
from~\cite{seminal} as well as ill-typed OCaml programs we took from
the GRASShopper program verification tool~\cite{grasshopper}.  The
student benchmark suite consists of OCaml programs written by students
that were new to OCaml.  We took the $356$ programs from the benchmark
suite that are ill-typed. Most of these programs exhibit type mismatch
errors. Only few of programs have trivial type errors such as calling
a function with too many arguments or assigning a non-mutable field of
a record.  The other programs in the benchmark suite that we did not
consider do not exhibit type errors, but errors that are inherently
localized, such as the use of an unbounded value or constructor.  The
size of these programs is limited; the largest example has $397$ lines
of code.

Since we lacked an appropriate corpus of larger
ill-typed user written programs, we generated ill-typed programs from
the source code of the GRASShopper tool~\cite{grasshopper}. We chose
GRASShopper because it contains non-trivial code that mostly falls
into the OCaml fragment supported by EasyOCaml. For our
experiments, we took several modules from the GRASShopper source
code and put them together into four programs of $1000$, $1500$,
$2000$, and $2500$ lines of code, respectively. These modules include
the core data structures for representing the abstract syntax trees of
programs and specification logics, as well as complex utility
functions that operate on these data structures. We included comments
when counting the number of program lines. However, comments were
generally scars. The largest program with $2500$ lines comprised 282
top-level $\Let$ definitions and 567 $\Let$ definitions in total.  We
then introduced separately five distinct type errors to each program,
obtaining a new benchmarks suite of $20$ programs in total.  We
introduced common type mismatch errors such as calling a function or
passing an argument with an incompatible type.


All of our timing experiments were conducted on a $3.60$GHz \texttt{Intel(R) Xeon(R)} 
machine with $16$GBs of RAM.

\begin{figure}[h]
    \centering
    \includegraphics[width=0.35\textwidth, width=0.35\textwidth, angle=-90]{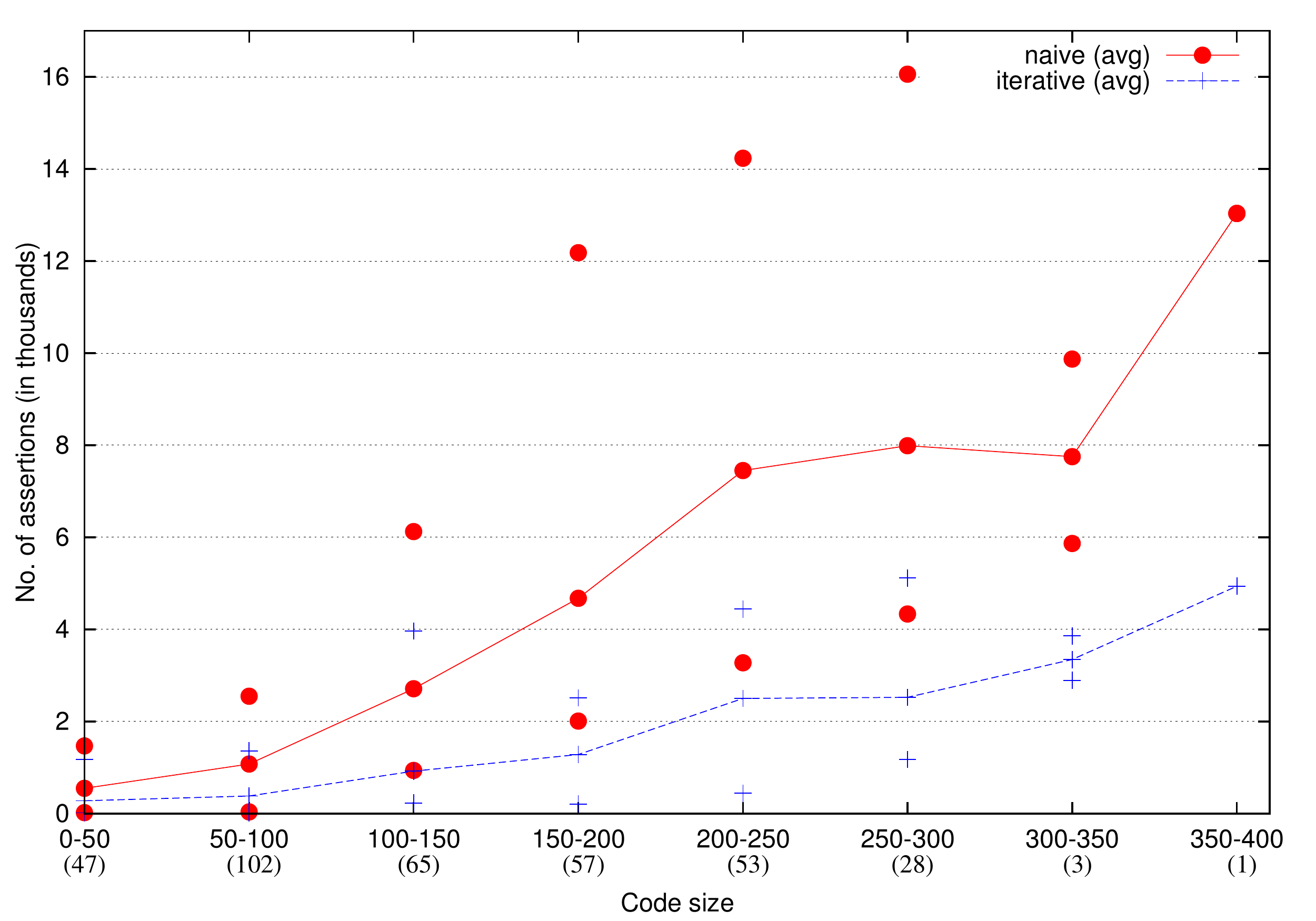}
    \caption{Maximum, average, and minimum number of typing assertions for computing a minimum error source by naive and iterative approach}
    \label{fig:num-constraints}
\end{figure}

\begin{figure}[t]
    \centering
    \includegraphics[width=0.35\textwidth, width=0.35\textwidth, angle=-90]{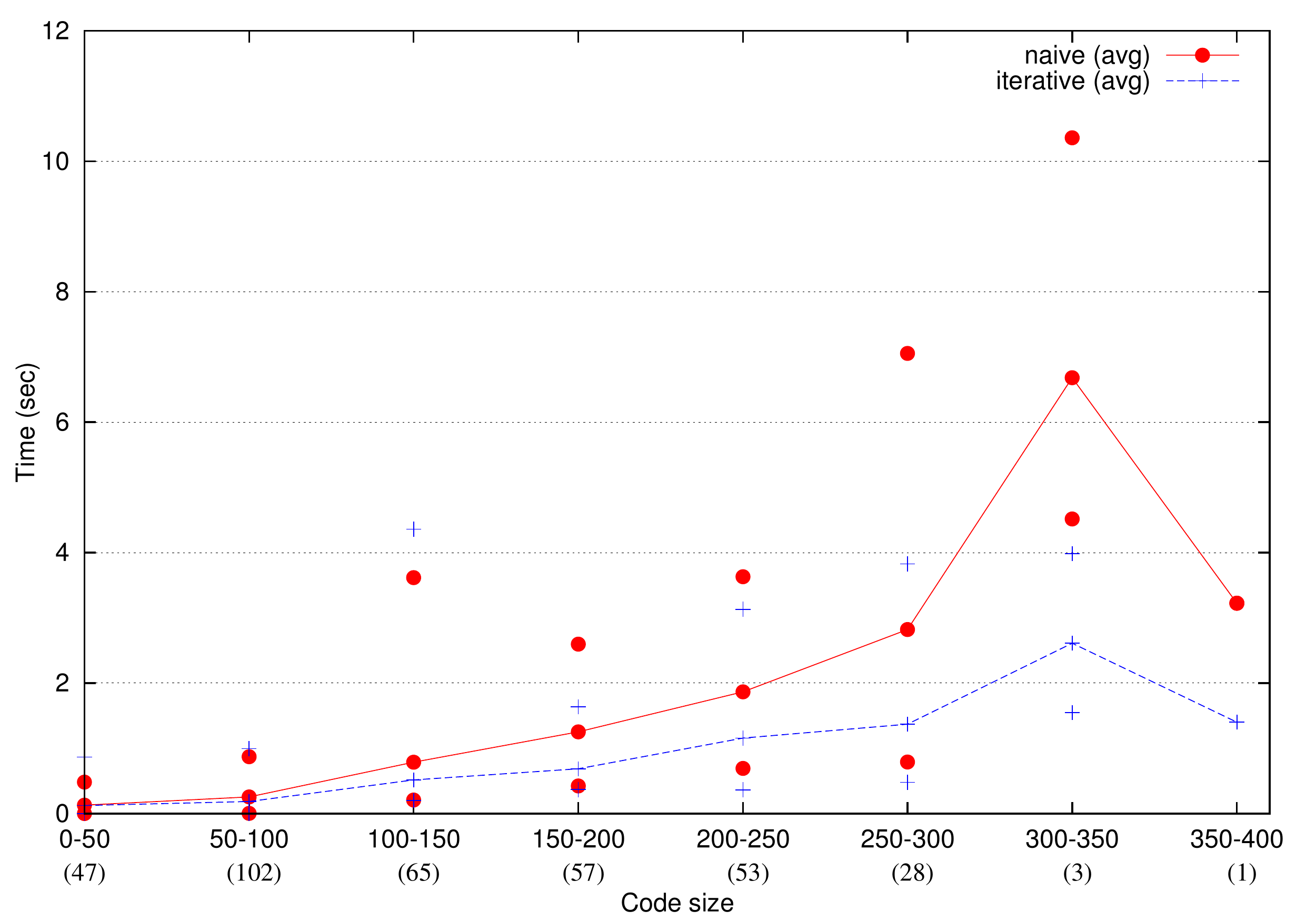}
    \caption{Maximum, average, and minimum execution times for computing a minimum error source by naive and iterative approach}
    \label{fig:execution-times}
\end{figure}

\paragraph{Student benchmarks.} In our first experiment,
we collected statistics for finding a single minimum error source in
the the student benchmarks with our iterative algorithm
and the naive algorithm from~\cite{minerrors}.
We measured the number of typing constraints generated
(Fig.~\ref{fig:num-constraints}),
the execution times (Fig.~\ref{fig:execution-times}), and
the number of expansions and iterations taken by our algorithm
(Table~\ref{tab:expansions}).
The benchmark suite of $356$ programs is broken into $8$ groups according to the number 
of lines of code in the benchmark.
The first group includes programs consisting of $0$ and $50$ lines of code,
the second group includes programs of size $50$ to $100$, and so on.

Figure~\ref{fig:num-constraints} shows the statistics for the total number of
generated typing assertions. By typing assertions we mean logical assertions,
encoding the typing constraints, that we pass to the weighted MaxSMT solver. The
number of typing assertions roughly corresponds to the sum of the total number of locations,
constraints attached to each location due to copying, and the number of well typed
\texttt{let} definitions. All $8$ groups of programs are shown on the \texttt{x} axis in Figure~\ref{fig:num-constraints}.
The numbers in parenthesis indicate the number of programs in each group.
For each group and each approach (naive and iterative), we plot the maximum,
minimum and average number of typing assertions.
To show the general trend for how both approaches are scaling, 
lines have been drawn between the averages for each group.
(All of the figures in this section follow this pattern.)
As can be seen, our algorithm reduces the total number of generated typing assertions.
This number grows exponentially with the size of the program for the naive approach.
With our approach, this number seems to grow at a much slower rate since it does not
expand every usage of a \texttt{let} variable unless necessary.
These results make us cautiously optimistic that the number of assertions the iterative
approach expands will be polynomial in practice.
Note that the total number of typing assertions produced by our algorithm is the
one that is generated in the last iteration of the algorithm.   

The statistics for execution times are shown in Figure~\ref{fig:execution-times}.
The iterative algorithm is consistently faster than the naive solution.
We believe this to be a direct consequence of the fact that our algorithm generates a
substantially smaller number of typing constraints.
The difference in execution times between our algorithm and the naive approach
increases with the size of the input program. Note that the total times shown
are collected across all iterations.

We also measured the statistics on the number of iterations and
expansions taken by our algorithm.  The number of expansions
corresponds to the total number of usage locations of \texttt{let}
variables that have been expanded in the last iteration of our
algorithm.  The results, shown in Table~\ref{tab:expansions}, indicate
that the total number of iterations required does not substantially
change with the input size.  We hypothesize that this is due to the
fact that type errors are usually tied only to a small portion of the
input program, whereas the rest of the program is not relevant to the
error.

\begin{table}[t]
	\center
        \begin{tabular}{c|c|c|c||c|c|c|}
                \cline{2-7} & \multicolumn{3}{c}{\textbf{iterations}} & \multicolumn{3}{c|}{\textbf{expansions}} \\
                \cline{2-7} & min & avg & max & min & avg & max \\ \hline
		\multicolumn{1}{|c||}{0-50} & 0 & 0.49 & 2 & 0 & 1.7 & 11\\ \hline 
		\multicolumn{1}{|c||}{50-100} & 0 & 0.29 & 3 & 0 & 0.88 & 13\\ \hline 
		\multicolumn{1}{|c||}{100-150} & 0 & 0.49 & 4 & 0 & 1.37 & 32\\ \hline 
		\multicolumn{1}{|c||}{150-200} & 0 & 0.44 & 3 & 0 & 1.82 & 19\\ \hline 
		\multicolumn{1}{|c||}{200-250} & 0 & 0.49 & 2 & 0 & 3.11 & 30\\ \hline 
		\multicolumn{1}{|c||}{250-300} & 0 & 0.36 & 2 & 0 & 6.04 & 45\\ \hline 
		\multicolumn{1}{|c||}{300-350} & 0 & 0.67 & 2 & 0 & 3.33 & 10\\ \hline 
		\multicolumn{1}{|c||}{350-400} & 0 & 0 & 0 & 0 & 0 & 0\\ \hline 
        \end{tabular}
    \caption{Statistics for the number of expansions and iterations when computing a single minimum error source}
	\label{tab:expansions}
\end{table}

It is worth noting that both the naive and iterative algorithm compute
single error sources.  The algorithms may compute different solutions
for the same input since the fixed cost function does not enforce
unique solutions.\footnote{ Both approaches are complete and would
  compute identical solutions for the all error sources
  problem~\cite{minerrors}.  } The iterative algorithm does not
attempt to find a minimum error source in the least number of
iterations possible, but rather it expands \texttt{let} definitions
on-demand as they occur in the computed error sources.  This means
that the algorithm sometimes continues expanding \texttt{let}
definitions even though there exists a proper minimum error source for
the current expansion. In our future work, we plan to consider how to
enforce the search algorithm so that it first finds those minimum
error sources that require less iterations and expansions.

\paragraph{GRASShopper benchmarks.}
We repeated the previous experiments on the generated GRASShopper benchmarks.
The benchmarks are grouped by code size.
There are four groups of five programs corresponding to programs
with 1000, 1500, 2000, and 2500 lines.

\begin{figure}[p]
    \centering
    \includegraphics[width=0.35\textwidth, width=0.35\textwidth, angle=-90]{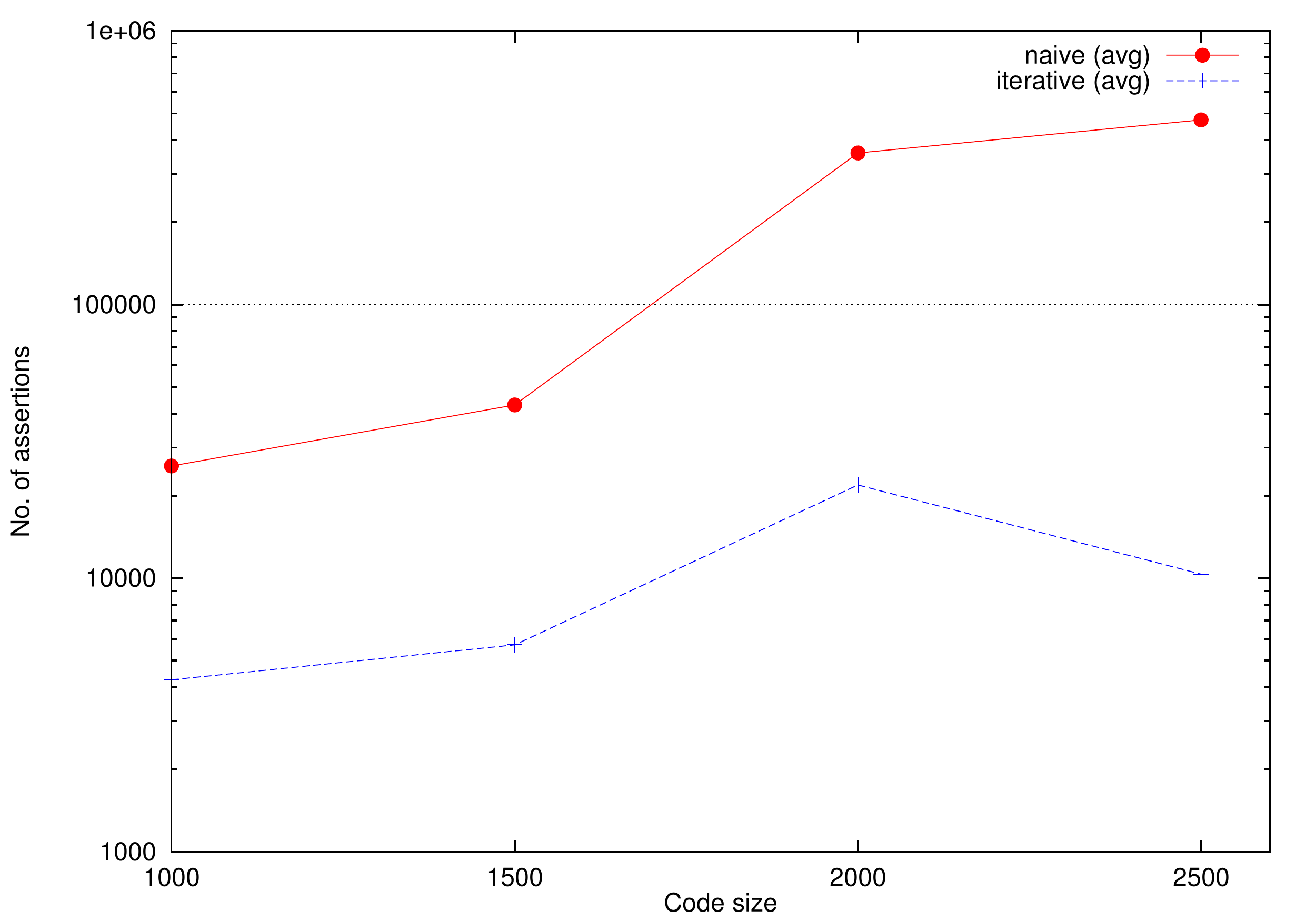}
    \caption{Maximum, average, and minimum number of typing assertions for computing a minimum error source by naive and iterative approach for larger programs} 
    \label{fig:num-constraints-larger}
\end{figure}

\begin{figure}[p]
    \centering
    \includegraphics[width=0.35\textwidth, width=0.35\textwidth, angle=-90]{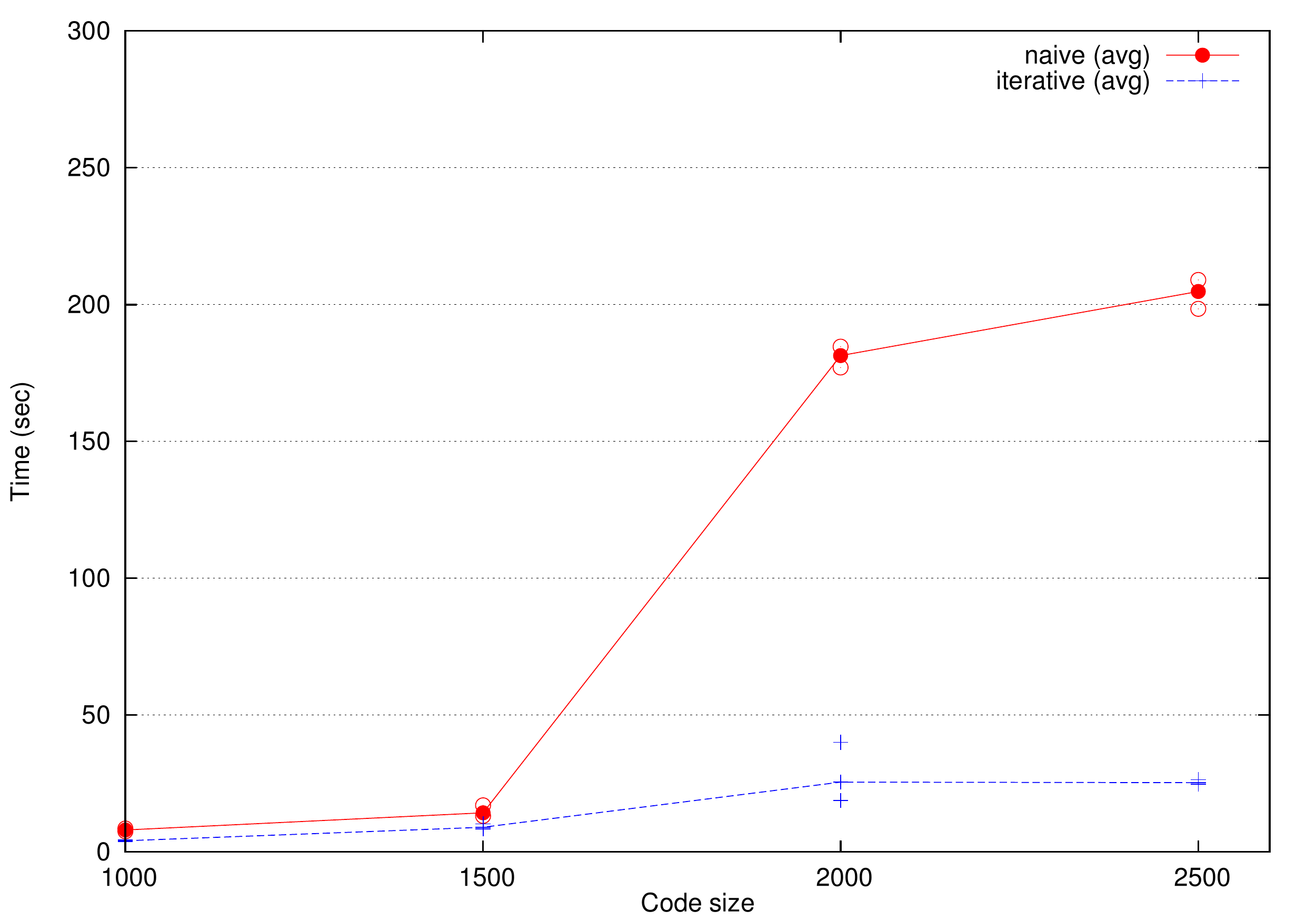}
    \caption{Maximum, average, and minimum execution times for computing a minimum error source by naive and iterative approach for larger programs}
    \label{fig:execution-times-larger}
\end{figure}

Figure~\ref{fig:num-constraints-larger} shows the total number of generated
typing assertions subject to the code size.
This figure follows the conventions of Fig.~\ref{fig:num-constraints}
except that the number of constraints is given on a logarithmic scale.\footnote{
  The minimum, maximum, and average points are plotted in
  Figures~\ref{fig:num-constraints-larger} and~\ref{fig:execution-times-larger}
  for each group and algorithm,
  but these are relatively close to each other and hence visually mostly indistinguishable.
}
The total number of assertions generated by our algorithm is consistently
an order of magnitude smaller than when using the naive approach.
The naive approach expands all \texttt{let} defined variables where the iterative approach
expands only those \texttt{let} definitions that are needed to find the minimum
error source.
Consequently, the times taken by our algorithm to compute a minimum error
source are smaller than when using the naive one, as shown in
Figure~\ref{fig:execution-times-larger}.
Beside solving a larger weighed MaxSMT instance, the naive approach also
has to spend more time generating typing assertions than our iterative
algorithm. 
Finally, Table~\ref{tab:expansions-larger} shows the statistics on the number
of iterations and expansion our algorithm made while computing the minimum
error source. Again, the total number of iterations appears to be
independent of the size of the input program.

\begin{table}[p]
	\center
  \begin{tabular}{c|c|c|c||c|c|c|}
                \cline{2-7} & \multicolumn{3}{c}{\textbf{iterations}} & \multicolumn{3}{c|}{\textbf{expansions}} \\
                \cline{2-7} & min & avg & max & min & avg & max \\ \hline
		\multicolumn{1}{|c||}{1000} & 0 & 0.2 & 1 & 0 & 0.2 & 1\\ \hline 
		\multicolumn{1}{|c||}{1500} & 0 & 0.4 & 2 & 0 & 2.8 & 14\\ \hline 
		\multicolumn{1}{|c||}{2000} & 0 & 0.6 & 2 & 0 & 53.8 & 210\\ \hline 
		\multicolumn{1}{|c||}{2500} & 0 & 0.2 & 1 & 0 & 3 & 15\\ \hline 
  \end{tabular}
  \caption{Statistics for the number of expansions and iterations when computing a single minimum error source for larger programs}
	\label{tab:expansions-larger}
\end{table}

\paragraph{Comparison to other tools.}
Our algorithm also outperforms the approach by Myers and Zhang~\cite{myers} in
terms of speed on the same student benchmarks. While our algorithm ran always
under $5$ seconds, their algorithm took over $80$ seconds for some programs.
We also ran their tool SHErrLoc~\cite{sherrloc} on one of our GRASSHopper benchmark
programs of $2000$ lines of code. After approximately $3$ minutes, their tool
ran out of memory. We believe this is due to the exponential explosion in
the number of typing constraints due to polymorphism. For that particular
program, the total number of typing constraints their tool generated was roughly
$200,000$. On the other hand, their tool shows high precision in correctly
pinpointing the actual source of type errors. These results nicely exemplify 
the nature of type error localization. In order to solve the problem of
producing high quality type error reports, one needs to consider the whole 
typing data. However, the size of that data can be impractically large, making 
the generation of type error reports slow to the point of being not usable. 
One benefit of our approach is that these two problems can be studied independently. 
In this work, we focused on the second problem, i.e., how to make the search for
high-quality type error sources practically fast.


\section{Conclusion}

We have presented a new algorithm that efficiently finds optimal type
error sources subject to generic usefulness criteria. The algorithm
uses SMT techniques to deal with the large search space of potential
error sources, and principal types to abstract the typing constraints
of polymorphic functions. The principal types are lazily expanded to
the actual typing constraints whenever a candidate error source
involves a polymorphic function. This technique avoids the
exponential-time behavior that is inherent to type checking in the
presence of polymorphic functions and still guarantees the optimality
of the computed type error sources. We experimentally showed that our
algorithm scales to programs of realistic size. To our knowledge, this
is the first type error localization algorithm that guarantees optimal
solutions and is fast enough to be usable in practice.


\paragraph{Acknowledgments}
This work was in part supported by the \href{http://www.nsf.gov/}{National Science Foundation} under grant CCF-1350574 and the \href{http://erc.europa.eu/}{European Research Council} under the
European Union's Seventh Framework Programme (FP/2007-2013) / ERC Grant
Agreement nr.~306595 \href{http://stator.imag.fr/}{\mbox{``STATOR''}}.


\balance
\bibliographystyle{abbrv}
\bibliography{main}

\end{document}